\documentstyle[multicol,aps,epsfig,graphicx,amstex]{revtex}
\newcommand{\la}{\langle}
\newcommand{\ra}{\rangle}

\newcommand{\mA}{{\mathcal A}}
\newcommand{\mB}{{\mathcal B}}
\newcommand{\mE}{{\mathcal E}}

\renewcommand{\Im}{{\mathrm Im}}
\renewcommand{\Re}{{\mathrm Re}}
\newcommand{\Tr}{{\mathrm Tr}}

\begin{document}
\title{Nonclassical statistics of intracavity coupled $\chi^{(2)}$
  waveguides: the quantum optical dimer}

\author{M. Bache$^{1,2}$, Yu. B. Gaididei$^1$ and P. L. Christiansen$^1$} 
\address{ 1)
  Informatics and Mathematical Modelling, Technical University of
  Denmark,
  DK-2800 Lyngby, Denmark.\\
  2) Optics and Fluid Dynamics Department, Ris{\o} National
  Laboratory, Postbox 49, DK-4000 Roskilde, Denmark.
  \\
}

 \date{\today, resubmitted version}
\maketitle
\begin{abstract}
  
  A model is proposed where two $\chi^{(2)}$ nonlinear waveguides are
  contained in a cavity suited for second-harmonic generation. The
  evanescent wave coupling between the waveguides is considered as
  weak, and the interplay between this coupling and the nonlinear
  interaction within the waveguides gives rise to quantum violations
  of the classical limit. These violations are particularly strong
  when two instabilities are competing, where twin-beam behavior is
  found as almost complete noise suppression in the difference of the
  fundamental intensities.  Moreover, close to bistable transitions
  perfect twin-beam correlations are seen in the sum of the
  fundamental intensities, and also the self-pulsing instability as
  well as the transition from symmetric to asymmetric states display
  nonclassical twin-beam correlations of both fundamental and
  second-harmonic intensities.  The results are based on the full
  quantum Langevin equations derived from the Hamiltonian and
  including cavity damping effects. The intensity correlations of the
  output fields are calculated semi-analytically using a linearized
  version of the Langevin equations derived through the positive-$P$
  representation.  Confirmation of the analytical results are obtained
  by numerical simulations of the nonlinear Langevin equations derived
  using the truncated Wigner representation.
 
\end{abstract}


\begin{multicols}{2}

\section{Introduction}
\label{sec:intro}

The $\chi^{(2)}$ nonlinear materials have been the subject of various
investigations in recent years. Using a cavity setup the weak
nonlinearities can be resonantly amplified, and complex spatiotemporal
behavior has been predicted from a classical point of view
\cite{lugiato:1999}. Moreover, due to the quantum fluctuations of
light many interesting non-classical effects have been reported, such
as squeezed light \cite{kimble:1987} and sub-Poissonian light
\cite{davidovich:1996}, both theoretically
\cite{drummond:1981,collett:1985} and experimentally
\cite{slusher:1985,pereira:1988}. The interplay between the classical
spatial instabilities and the quantum fluctuations in the system has
been investigated intensively lately \cite{lugiato:1999}, a study
devoted to characterizing the mode interaction on the quantum level.

We consider the case of second-harmonic generation (SHG), where the
photons of the pump field (fundamental photons) are up-converted in
pairs to second-harmonic photons of the double frequency.  The model
we propose in this paper consists of two quadratically nonlinear
waveguides placed in a cavity that resonates both the fundamental and
second-harmonic, and we take linear coupling between the waveguides
into account. In a sense this is the simplest mode coupling model
obtainable. The question is how the coupling between the waveguides
affects the cavity dynamics, and in particular we shall focus on the
nonclassical behavior of the system. 

The name proposed for this model, the quantum optical dimer,
originates from the numerous investigations made about discrete site
coupling in various systems, such as condensed matter physics and
biology, see Ref.\cite{scott:1990} for a general treatment of discrete
systems. Thus, the name dimer implies that coupling between two
discrete sites are being taken into account. For a single waveguide
(or site) we shall use the name monomer, a case corresponding here to
a bulk nonlinear medium and neglecting diffraction.

It has been shown that in the SHG quantum optical monomer excellent
squeezing of the output fields is possible close to a self-pulsing
instability \cite{collett:1985}, and nonclassical effects in SHG have
been verified experimentally \cite{pereira:1988}, and even shown to
persist above the threshold of the instability \cite{pettiaux:1991}.
Also in the presence of diffraction strong correlations exist between
different spatial modes in the presence of a spatial instability
\cite{bache:2002b}, including strong correlations between the
fundamental field and the generated second-harmonic field as well as
spatial multimode nonclassical light.

As we shall show the quantum optical dimer also displays strong
nonclassical correlations in the intensity correlations, and that the
linear coupling across the waveguides plays a decisive role.  The
model has three types of transitions, bistability, self-pulsing and a
symmetric to asymmetric transition. It is remarkable that particularly
strong nonclassical correlations are observed when two of these
instabilities compete. Specifically, when taken close to a
self-pulsing or bistable regime the symmetric to asymmetric transition
has nearly perfect \textit{twin-beam} behavior, so the difference of
the fundamental intensities displays almost no fluctuations. The
twin-beam correlations were first shown in the optical parametric
oscillator (OPO) \cite{reynaud:1987,heidmann:1987}, where the signal
and idler photons of the twin-fields are generated simultaneously from
the pump field, and the intensity difference shows correlations below
the classical limit.  However, the twin-beam effect observed in the
dimer originates from photons created in different waveguides with
only the coupling to link them. Thus, the photons are strictly
speaking not twins, but merely ``brothers''. The twin-beam effect is
also observed near bistable turning points where complete noise
suppression is observed in the sum of the fundamental intensities, the
strongest violations occurring in the limit where the fundamental
input coupling loss rate is much larger than the second-harmonic one.
That bistability gives rise to highly nonclassical effects turns out
to also hold for the SHG monomer, and has to our knowledge not been
observed before; usually the self-pulsing transition has been used to
observe violations of the classical limit, which we also observe in
the dimer. The bistable transition has previously been observed to
produce nonclassical states in other systems, such as dispersive and
absorptive optical bistability \cite{collett:1985,lugiato:1982} and
Raman lasers \cite{eschmann:1999}.

A closely related optical model is spatially coupled lasers
\cite{fabiny:1993}, where a single laser medium is pumped by two beams
spatially separated. Waveguiding is achieved by thermal lensing, in
which the temperature dependent refractive index of the laser medium
creates a guiding effect, and the coupling strength is controlled by
the distance between the pump profiles. The quantum noise induced
correlations in these systems have not yet been reported to beat the
standard quantum limit when Kerr type nonlinearities are considered
\cite{serrat:2001,serrat:2002}, except when the coupling arises solely
due to initially correlated noise terms of the pumps
\cite{khoury:1999}.

The cavityless setup of coupled $\chi^{(2)}$ waveguides has previously
been investigated, both from a classical and a quantum mechanical
point of view. In the classical model of waveguide arrays, the focus
of attention has been on soliton behavior originating from the
coupling \cite{peschel:1997}, whereas the cavityless dimer was shown
to produce chaotic states away from the integrable limit (where
second-harmonic coupling is neglected) \cite{bang:1997}.  The quantum
behavior of the cavityless dimer has been investigated by the group
of Perina \textit{et al.}  (for a review see Ref. \cite{perina:2000})
giving the name ``nonlinear coupler'' to the model. They have
investigated both co- and counter-propagating input fields in
parametric oscillation and, \textit{e.g.}, the transfer of quantum
states from one waveguide to the other.

The model presented here is also closely related to the dynamics of
coupled atomic and molecular Bose-Einstein condensates (BECs)
\cite{wynar:2000,heinzen:2000}, where the photoassociation of an
atomic condensate may produce a molecular condensate with an
atom-molecule interaction that is reminiscent of the interaction
between the fundamental and second-harmonic photons in SHG
\cite{drummond:1998}. The opposite process where the photo
dissociation of a molecular BEC creates an atomic BEC has been shown
to produce squeezed states \cite{poulsen:2001}, a model which has the
quantum optical equivalent in the OPO. If an analogy should be drawn
between the quantum optical dimer presented here and BEC it would
consist of placing two such coupled molecular-atomic BECs in separate
quantum wells. Thus, evanescent tunneling of the wave functions
between the wells would introduce the dimer coupling, similar to what
is done in Ref.  \cite{milburn:1997} for a normal BEC.

We should finally stress that the cavity setup discussed in the
present work gives rise to two major differences to the work in
cavityless waveguides as well as for the BEC. First of all the cavity
introduces losses in the model through the input mirror, and secondly
external pump fields appear in the equations acting as forcing terms.

The paper is structured as follows. In Sec.~\ref{sec:model} the model
is introduced, and the stochastic Langevin equations are derived from
the full boson operator Hamiltonian. Also, we discuss the allowed
values of the coupling constants. In Sec.~\ref{sec:linear-stability}
the linear stability of the Langevin equations are investigated, and
the bifurcation scenario of the model is discussed. In
Sec.~\ref{sec:squeezing-analysis} we discuss the framework for the
two-time photon number correlations of the output fields, and the
semi-analytical spectral variances are derived in the linearized
limit. Section~\ref{sec:Results} is devoted to the results of the
analytical calculations as well as the numerical simulations. A
summary is made in Sec.~\ref{sec:summary} where we also discuss the
results obtained.  Appendix~\ref{sec:Fokk-Planck-descr} shows details
about the derivation of the quasi-probability distribution equations
used to connect the master equation for the quantum Hamiltonian with
the classical looking stochastic Langevin equations. The numerical
methods are discussed in App.~\ref{sec:num}.

\begin{figure}
\includegraphics[width=8cm]{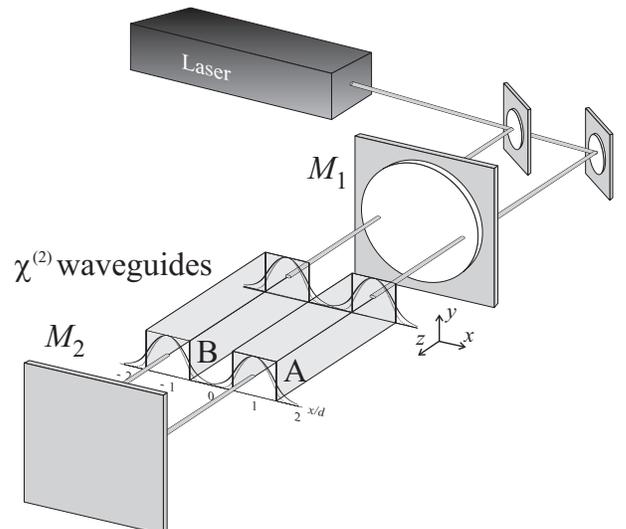}
\caption{The setup. Two nonlinear waveguides A and B inside a cavity
  pumped by a classical field.}
\label{fig:setup}
\end{figure}

\section{The model}
\label{sec:model}

We consider the setup shown in Fig.~\ref{fig:setup}.  Two $\chi^{(2)}$
nonlinear waveguides are contained in a cavity with a high reflection
input mirror $M_1$ and a fully reflecting mirror $M_2$ at the other
end. The cavity is pumped at the frequency $\omega_1$ and through the
nonlinear interaction in the waveguides photons of the frequency
$\omega_2=2\omega_1$ are generated; this is the process of
second-harmonic generation. The cavity supports a discrete number of
longitudinal modes, and we will consider the case where only two of
these modes are relevant, namely the mode $\omega_{1,{\mathrm cav}}$
closest to the fundamental-harmonic (FH) frequency and
$\omega_{2,{\mathrm cav}}$ closest to the second-harmonic (SH)
frequency.  Using the mean field approximation the $z$-direction, in
which the pump beam propagates, is averaged out. This approach is
justified as long as the losses and detunings are small, and we
furthermore assume perfect phase matching.  The waveguiding implies
that diffraction in the transverse plane may be neglected. Let $\hat
A_1(t)$ and $\hat B_1(t)$ ($\hat A_2(t)$ and $\hat B_2(t)$) denote the
FH (SH) intracavity boson operators of waveguide A and B,
respectively. They are normalized so they obey the following equal
time commutation relations
\begin{equation}
  \label{eq:Acomm}
  [\hat O_i (t),\hat O_j^{\dagger}(t)]=\delta_{ij},
\quad i,j=1,2, \quad \hat O=\hat A,\hat B,
\end{equation}
while $[\hat A_j(t),B^\dagger_j(t)]=0$. The system is modelled through
the Hamiltonian
\begin{equation}
  \label{eq:H-dimer}
  \hat H=\hat H^{\mathrm sys}_A+\hat H^{\mathrm sys}_{B}+\hat H_{AB}, 
\end{equation}
where the system Hamiltonians in the frame rotating with the pump
frequency are given by
\begin{eqnarray}
    \hat H^{\mathrm sys}_O&=&-\hbar\delta_1 \hat O_1^{\dagger}
  \hat O_1-\hbar\delta_2\hat O_2^{\dagger} \hat O_2
+\frac{i \hbar \kappa}{2}({\hat O^{\dagger 2}_1} \hat
  O_2-\hat O_1^{2} \hat O_2^{\dagger})
\nonumber \\
  &&+i\hbar ({\mathcal E}_{{\mathrm p},O} \hat O_1^\dagger  -{\mathcal
  E}_{{\mathrm p},O}^*\hat O_1),\quad O=A,B.
\end{eqnarray}
The detunings from the cavity resonances are given by
$\delta_j=\omega_j-\omega_{j,\mathrm cav}$, $\kappa$ is proportional
to the $\chi^{(2)}$ nonlinearity and ${\mathcal E}_{{\mathrm p},O}$
are the external pump fields at the FH frequency of the individual
waveguides \cite{opo}, here treated as classical fields. The coupling
between the waveguides is modelled as overlapping tails of evanescent
waves so it may be assumed weak, implying we can describe it as a
linear process
\begin{equation}
  \label{eq:H_AB}
    \hat H_{AB}=\hbar J_1(\hat A_1\hat B_1^\dagger+\hat B_1\hat A_1^\dagger)+
  \hbar J_2(\hat A_2\hat B_2^\dagger+\hat B_2\hat A_2^\dagger).
\end{equation}
$J_1$ and $J_2$ are the cross-waveguide coupling parameters of the FH
and SH, respectively. The time evolution of the reduced system density
matrix operator $\hat \rho$ in the Schr\"odinger picture is then given
by the master equation \cite{walls:1994,carmichael}
\begin{equation}
  \label{eq:ME}
  \frac{\partial\hat \rho}{\partial
  t}=-\frac{i}{\hbar}[\hat H,\hat\rho]+(\hat L_{1,A}+ 
  \hat L_{2,A}+ \hat L_{1,B}+\hat L_{2,B})\hat\rho.
\end{equation}
The continuum of modes outside the cavity is modelled as a heat bath
in thermal equilibrium, and the coupling to these modes has been
included through the Liouvillian terms
\begin{eqnarray}
  \label{eq:Liouvillian}
  \hat L_{j,O}\hat\rho&=&\gamma_j\left([\hat O_j,\hat\rho \hat
    O_j^\dagger]+ [\hat O_j \hat\rho,\hat O_j^\dagger]\right)
  \nonumber\\&&+
  \gamma_j\bar n^{\mathrm th}_{j} \left([\hat O_j\hat\rho,\hat
    O_j^\dagger]+[\hat 
    O_j^\dagger, \hat\rho\hat O_j]\right).
\end{eqnarray}
These terms describe the losses of the fields through photons escaping
the cavity, and simultaneously they model fluctuations entering the
cavity through the input mirror, a consequence of the
dissipation-fluctuation theorem \cite{mandel:1995}. The loss rates of
the input coupling mirror are given by $\gamma_j$, whereas the terms
$\bar n^{\mathrm th}_{j}=(e^{\hbar\omega_j/k_BT}-1)^{-1}$ are the mean
number of thermal quanta in the external bath modes at $\omega_j$. We
shall here neglect thermal fluctuations by setting the bath
temperature $T=0$ yielding $\bar n^{\mathrm th}_{j}=0$. First of all
this is a good approximation for optical systems since here $\hbar
\omega\gg k_BT$, and secondly we may hereby focus on behavior solely
due to the inherent quantum fluctuations of light.

The master equation~(\ref{eq:ME}) is difficult to solve as it is,
therefore we apply the now standard technique of expanding the density
matrix in a basis of coherent states weighted by a quasi-probability
distribution (QPD). The details of this quantum-to-classical
description are given in App.~\ref{sec:Fokk-Planck-descr}, and the
result is a partial differential equation of the QPD. This QPD
equation depends on the choice of ordering of the corresponding
quantum mechanical averages. Equation~(\ref{eq:FPE-chi2-dimer}) is the
QPD equation using the positive-$P$ distribution giving normally ordered
averages, which we will use for the linearized analysis. For the
numerical implementation the Wigner distribution is used to obtain
Eq.~(\ref{eq:c-number-Wigner-dimer}), in which symmetric averages are
calculated.

If the QPD equations~(\ref{eq:FPE-chi2-dimer})
and~(\ref{eq:c-number-Wigner-dimer}) are on the Fokker-Planck
form~(\ref{eq:FPE-general1}), an equivalent set of stochastic Langevin
equations~(\ref{eq:lang-ito-general-simple}) can be found to by using
Ito rules of stochastic integration \cite{ito-stratonovich}.  For the
Wigner QPD equation~(\ref{eq:c-number-Wigner-dimer}) this is not the
case because of the third order terms, however these terms, which have
been shown to model quantum jump processes \cite{kinsler:1991}, are
generally neglected and the resulting Fokker-Planck equation turns out
to be a good approximation to the original problem. Using this
approximation the normalized Langevin equations for the Wigner QPD
equation are
\begin{mathletters}\label{eq:langevin-W}
\begin{eqnarray}
    \dot{A}_1&=&(-1 +i\Delta_1)A_1
  + A_1^*A_2 -iJ_1B_1 + \sqrt{2}A_{{\mathrm in},1}(t)
\\
  \dot{A}_2&=&(-\gamma +i\Delta_2)A_2
  - \frac{1}{2}A_1^2-iJ_2B_2 + \sqrt{2\gamma}A_{{\mathrm in},2}(t)
\\
  \dot{ B}_1&=&(-1 +i\Delta_1)B_1
  + B_1^*B_2 
  -iJ_1A_1 + \sqrt{2}B_{{\mathrm in},1}(t)
\\
  \dot{B}_2&=&(-\gamma +i\Delta_2)B_2
  - \frac{1}{2}B_1^2-iJ_2A_2 +
\sqrt{2\gamma}B_{{\mathrm in},2}(t),
\end{eqnarray}
\end{mathletters}
where the dot denotes derivation with respect to time. The fields
$\{A_j,A_j^*\}$ and $\{B_j,B_j^*\}$ are normalized equivalent
$c$-numbers to the operators $\{\hat A_j,\hat A_j^\dagger\}$ and
$\{\hat B_j,\hat B_j^\dagger\}$. The input fields are describing the
pump field entering the cavity through the input mirror as well as the
noise coupled in here according to the Liouvillian
terms~(\ref{eq:Liouvillian})
\begin{mathletters}\label{eq:noise-W}
  \begin{eqnarray}
    \label{eq:FH-noise-W}
    F_{{\mathrm in},1}(t)=\frac{E}{\sqrt{2}}+\xi_{F_1}(t), \quad
    F_{{\mathrm in},2}(t)=\xi_{F_2}(t) \\
\la \xi_{F_i}^*(t)\xi_{F_j}(t')\ra=\delta_{ij}\frac{\delta(t-t')}{2n_s},
  \end{eqnarray}
\end{mathletters}
with $F=A,B$. All other correlations are zero. The positive-$P$ QPD
equation~(\ref{eq:FPE-chi2-dimer}) is on Fokker-Planck form so no
approximations are needed. The equivalent set of Langevin equations is
given by~(\ref{eq:langevin-W}) with $A_j^*\rightarrow A_j^\dagger$ and
$B_j^*\rightarrow B_j^\dagger$, as well as the equations for the
fields
\begin{mathletters}\label{eq:langevin-P}
\begin{eqnarray}
    \dot{A}_1^\dagger&=&(-1 -i\Delta_1)A_1^\dagger
  + A_1A_2^\dagger 
\nonumber\\&&
+iJ_1B_1^\dagger + \sqrt{2}A^\dagger_{{\mathrm in},1}(t)
\\
  \dot{A}_2^\dagger&=&(-\gamma -i\Delta_2)A_2^\dagger
  - \frac{1}{2}(A_1^\dagger)^2
\nonumber\\&&
+iJ_2B_2^\dagger +
  \sqrt{2\gamma}A^\dagger_{{\mathrm in},2}(t) 
\\
  \dot{ B}_1^\dagger&=&(-1 -i\Delta_1)B_1^\dagger
  + B_1B_2^\dagger 
\nonumber\\&&
  +iJ_1A_1^\dagger + \sqrt{2}B^\dagger_{{\mathrm in},1}(t)
\\
  \dot{B}^\dagger_2&=&(-\gamma -i\Delta_2)B_2^\dagger
  - \frac{1}{2}(B_1^\dagger)^2
\nonumber\\&&
+iJ_2A_2^\dagger +
\sqrt{2\gamma}B^\dagger_{{\mathrm in},2}(t).
\end{eqnarray}
\end{mathletters}
The fields $\{A_j,A_j^\dagger\}$ and $\{B_j,B_j^\dagger\}$ are normalized
equivalent $c$-numbers to the operators $\{\hat A_j,\hat A_j^\dagger\}$
and $\{\hat B_j,\hat B_j^\dagger\}$. The input fields for the
positive-$P$ Langevin equations are
\begin{mathletters}\label{eq:noise-P}
  \begin{eqnarray}
    \label{eq:FH-noise-p}
    F_{{\mathrm in},1}(t)=\frac{E}{\sqrt{2}}+\xi_{F_1}(t), \quad
    F^\dagger_{{\mathrm in},1}(t)=\frac{E}{\sqrt{2}}+\xi^\dagger_{F_1}(t)\\
    F_{{\mathrm in},2}(t)=0,\quad F^\dagger_{{\mathrm in},2}(t)=0 \\
\la \xi_{F_1}(t)\xi_{F_1}(t')\ra=\frac{F_2\delta(t-t')}{2n_s} \\
\la \xi^\dagger_{F_1}(t)\xi^\dagger_{F_1}(t')\ra= \frac{F_2^\dagger
    \delta(t-t')}{2n_s},
  \end{eqnarray}
\end{mathletters}
with $F=A,B$, and again all other correlations are zero. The doubling
of phase space associated with the positive-$P$ representation (see
App.~\ref{sec:Fokk-Planck-descr} for details) implies that
$\xi_j^\dagger$ is uncorrelated to $\xi_j$, and also that $A_j$ and
$A_j^\dagger$ are independent complex numbers and only on average is
$A_j^\dagger=A_j^*$.

The Langevin equations have been normalized by introducing the
dimensionless variables
\begin{mathletters}
\begin{eqnarray}
\tilde t=\gamma_1 t,\quad \gamma=\frac{\gamma_2}{\gamma_1}, \quad
  \Delta_j=\frac{\delta_j}{\gamma_1} \label{eq:space-time}\\
  A_j=\frac{\kappa}{\gamma_1}\alpha_j, \quad
  B_j=\frac{\kappa}{\gamma_1}\beta_j \\
A_{{\mathrm in},j}(\tilde t)=\frac{\kappa}{\gamma_1^{3/2}} \alpha_{{\mathrm
  in},j}(t), \quad  
B_{{\mathrm in},j}(\tilde t)=\frac{\kappa}{\gamma_1^{3/2}}\beta_{{\mathrm
  in},j}(t)
\label{eq:in-norm}
\\
\tilde \xi_j(\tilde t)=\frac{\kappa}{\gamma_1^{3/2}}\xi_j(t),
\quad  
E=\frac{\kappa}{\gamma_1^2}\mE_{\mathrm p}, \quad \tilde
  J_j=\frac{J_j}{\gamma_1},  \label{eq:J-norm}
\end{eqnarray}
\end{mathletters}
and the tildes have been dropped. The fields $\alpha_j$ and $\beta_j$
are the unscaled $c$-numbers, cf. App.~\ref{sec:Fokk-Planck-descr}.
We have furthermore introduced the dimensionless quantity
\begin{equation}
  \label{eq:ns}
  n_s=\kappa^2/\gamma_1^2.
\end{equation}
This parameter sets the level of the quantum noise, cf.
Eqs.~(\ref{eq:noise-W}) and (\ref{eq:noise-P}), and in the OPO it
represents the saturation photon number to trigger the parametric
oscillation.

For simplicity we have assumed real and equal pump rates in both
waveguides $\mE_{{\mathrm p},A}=\mE_{{\mathrm p},B}\equiv\mE_{{\mathrm
    p}}$. The consequence is that the same input mirrors as well as
intracavity paths are used for both waveguides, implying identical
detunings \cite{detuning} as well as losses for the FH fields and the
SH fields, respectively.

\begin{figure}
\includegraphics[width=8cm]{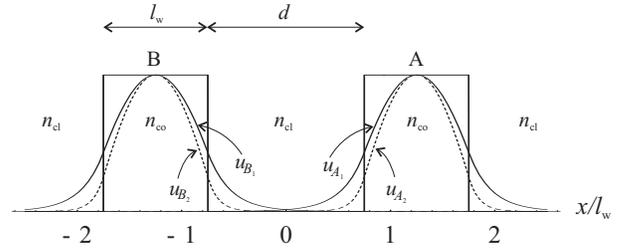}
\caption{Parallel planar waveguide setup shown in the
  $x$-direction. The width of the waveguides is $l_{\mathrm w}$ and
  the distance between them is $d$. The transverse distributions of
  the lowest order modes for a realistic setup are shown calculated
  using a step profile of the refractive index.}
\label{fig:waveguide}
\end{figure}

The coupling strengths between the waveguides are controlled by $J_1$
and $J_2$, and it is relevant to consider what values these may take.
Fig.~\ref{fig:waveguide} shows an instructive example, where we
consider symmetric step-index parallel planar waveguides with a core
(cladding) refractive index $n_{\mathrm co}$ ($n_{\mathrm cl}$). The
weakly guiding limit is assumed where $n_{\mathrm co}\simeq n_{\mathrm
  cl}$. Taking the FH field of waveguide A as example, the coupling
from waveguide B can be found by considering waveguide A in isolation
and taking the presence of waveguide B as a weak
perturbation. This approach assumes that the transverse
profile $u(x)$ and propagation constant $\beta$ of the modes in the
waveguides are left unchanged, and only the amplitude is modified by
the perturbation. The coupling constants of the propagation equations
of the waveguides are then found as \cite{saleh:1991}
\begin{equation}
  \label{eq:coupling-prop}
  J_{A_1B_1}=\frac{n_{\mathrm co}^2-n_{\mathrm
  cl}^2}{2}\frac{k_1^2} {\beta_{A_1}}\int_{-d/2-l_{\mathrm w}}^{-d/2} dx\;
  u_{A_1}(x)u_{B_1}(x), 
\end{equation}
where $k_1=2\pi/\lambda_1$ is the vacuum wavenumber of the FH, and the
mode profiles are assumed normalized so $\int_{-\infty}^\infty dx
u^2(x)=1$. Thus, Eq.~(\ref{eq:coupling-prop}) has the dimension
m$^{-1}$. Fig.~\ref{fig:waveguide} shows the lowest order modes of
the isolated waveguides as calculated for a realistic setup. If only
coupling between modes of same order is considered we have
$J_{A_1B_1}=J_{B_1A_1}\equiv J_1^{\mathrm prop}$. Applying the mean
field approach \cite{lodahl:1999} the coupling parameters of
Eqs.~(\ref{eq:langevin-W}) is then given by $J_j=J_j^{\mathrm prop}
L_{\mathrm cav}/\tau_1$, where $L_{\mathrm cav}$ is the length of the
cavity, and $\tau$ is the cavity round trip time. From
Eq.~(\ref{eq:J-norm}) the normalized coupling parameter is found
through $\gamma_1=T_1/(2\tau)$ where $T_1$ is the FH intensity
transmission efficiency of the input mirror, so we obtain
\cite{coupling}
\begin{equation}
  \label{eq:J-prop-conn}
  \tilde J_j=J_j^{\mathrm prop}\frac{2 L_{\mathrm cav}}{T_1}.
\end{equation}

As a result of these considerations we see that the SH coupling
parameter generally will be lower than the FH one. This is clear from
the calculated modes in Fig.~\ref{fig:waveguide}, where the SH modes
(dashed) decay faster than the FH modes (solid).  However, it is
impossible to generally say how much weaker and when the distance
between the waveguides is decreased the coupling parameters become
closer to each other. Finally, the actual values of $\tilde J_j$ are
highly sensitive to the specific setup. Not only in terms of waveguide
parameters (\textit{e.g.}, distance between guides, the modes in the
guides), but also on independent parameters (cavity length, input
transmission efficiency).  Using parameters from realistic setups
(similar to the cavity setup discussed in Ref.  \cite{lodahl:2001}) we
easily obtained normalized coupling parameters of $O(10^2)$, while
still preserving the assumptions of weak coupling as well as the
mean-field limit. Finally, when coupling between the lowest order
modes is considered we have $J_j>0$.

\section{Linear stability}
\label{sec:linear-stability}

In this section the Langevin equations derived previously are
linearized and the linear stability is investigated to obtain a
bifurcation scenario in the classical limit where noise is absent
($n_s\rightarrow \infty$). In this limit the Langevin equations from
the different representations give the same result, a natural
consequence from the fact that in the classical limit the operators
commute. Additionally, in Sec.~\ref{sec:squeezing-analysis} we are 
going to use the linearized equations with noise to derive analytical
results for the noise induced correlations. For this purpose it is
more convenient to use normally ordered intracavity averages, as will
be explained later, and this section will therefore only concern the
positive-$P$ Langevin equations.  The results of this section reveal
both symmetric and asymmetric modes in the two waveguides, as well as
bistable behavior and Hopf unstable solutions.

The linearization is particularly simple in the \textit{symmetric}
case.  Here the steady states in the waveguides are identical, so the
FH steady states in waveguide A and B are equal and equivalently for
the SH steady states. The symmetric steady states of the waveguides
can be found from the monomer equations, i.e. using the
results of Ref.~\cite{etrich:1997} and apply the substitution
$\Delta_j\rightarrow \Delta_j-J_j\equiv d_j$. In the symmetric case
the steady states are denoted $ \mA_j=\mB_j\equiv
\sqrt{\bar I_j}e^{i\phi_j}$ giving 
\begin{mathletters}
\begin{eqnarray}
  E^2&=&\bar I_1^2\frac{\bar I_1/4+(\gamma-d_1d_2)}{d_2^2+\gamma^2}+
  \bar I_1(d_1^2+1)\label{eq:E_I1} 
\\
\bar I_2&=&\bar I_1^2[4(d_2^2+\gamma^2)]^{-1}\label{eq:I2_I1}
\\
\phi_1&=&-{\mathrm Arg}\left(1-id_1+\bar I_1/[2(\gamma-id_2)]\right)
\\
\phi_2&=&-{\mathrm Arg}(-\gamma+id_2)+2\phi_1.
\end{eqnarray}
\end{mathletters}

We may linearize the positive-$P$ Langevin
equations~(\ref{eq:langevin-W}), (\ref{eq:langevin-P}) and
(\ref{eq:noise-P}) around the symmetric steady states $A_j=\Delta
A_j+\mA_j$ and $B_j=\Delta B_j+\mA_j$ \cite{classicalP} to get the
matrix equation
\begin{equation}
  \label{eq:int-pha-lin-mat}
  \Delta\dot{\mathbf w}={\mathbf A}\Delta{\mathbf w}
  +\frac{{\mathbf B}}{\sqrt{n_{s}}}{\mathbf n}(t),
\end{equation}
where $\Delta {\mathbf w}$ is a vector of fluctuations
\begin{eqnarray}
\nonumber  \Delta{\mathbf w}=
\left[
\Delta {A_1}, \Delta A_1^\dagger, \Delta {A_2}, \Delta
A_2^\dagger, \Delta {B_1}, \Delta B_1^\dagger, \Delta {B_2},
\Delta B_2^\dagger  
\right]^T,
\end{eqnarray} 
and ${\mathbf n}(t)$ is a vector of Gaussian white noise terms
correlated as 
\begin{equation}
  \label{eq:n(t)}
  \la n_{j}(t) n_{k}(t')\ra = \delta(t-t').
\end{equation}
The matrix ${\mathbf A}$ is block ordered into four $4\times 4$
matrices
\begin{equation}
  {\mathbf A}=
\begin{bmatrix}
    {\mathbf A}_{m} & {\mathbf A}_{x}\\
    {\mathbf A}_{x} & {\mathbf A}_{m}\\
\end{bmatrix},
\end{equation}
with the diagonal cross-coupling matrix
\begin{eqnarray}
  {\mathbf A}_{x}&=
  \mathrm{Diag} \left[
    \begin{array}{cccc}
      -iJ_1 & i J_1 & -iJ_2 & iJ_2 \\
    \end{array}\right],
\end{eqnarray}
and the monomer matrix is
\begin{eqnarray}
  {\mathbf A}_{m}&=
  \begin{bmatrix}
-1+i d_1 & {\mathcal A}_2 & {\mathcal A}_1^* & 0 \\
{\mathcal A}_2^* & -1-i d_1 & 0 & {\mathcal A}_1 \\
-{\mathcal A}_1 & 0 & -\gamma+id_2 & 0 \\
0 & -{\mathcal A}_1^* & 0 & -\gamma-id_2 \\    
  \end{bmatrix}.
\end{eqnarray}
The diffusion matrix is also diagonal
\begin{eqnarray}
  {\mathbf D}&=
  \mathrm{Diag}
  \begin{bmatrix}
      {\mathcal A}_2 & {\mathcal A}_2^* & 0 & 0 & {\mathcal A}_2 &
      {\mathcal A}_2^* & 0 & 0   \\
  \end{bmatrix},
\end{eqnarray}
and ${\mathbf D}={\mathbf B}^T{\mathbf B}$.

The classical stability of the system is found by solving the
eigenvalue problem $ {\mathbf A}{\mathbf v}=\lambda{\mathbf v}$, which
is done in \textsc{Mathematica}. The analysis is characterized by two
cases, either when one physical solution exists to the closed
problem~(\ref{eq:E_I1})-(\ref{eq:I2_I1}), or when the system is
bistable and three physical solutions exist (in this case each
solution must be analyzed individually). The stability of the steady
states may now either change with the critical eigenvalue
$\lambda_{j,c}$ having $\Im (\lambda_{j,c})=0$ at the critical pump
value $E_{\mathrm sym}$, which means that the symmetric state of the
system is no longer stable. When this happens a new state with
$\mA_j\neq \mB_j$ is stable instead, and the actual values of these
new steady states are not easily calculated. We will not address the
stability of the system beyond the asymmetric transition any further
in this paper, however the transition to the asymmetric state will be
used to look for nonclassical correlations.  The other possibility is
that the system changes stability with $\Im (\lambda_{j,c})\neq 0$ at
the critical pump value $E_{\mathrm SP}$, which corresponds to a Hopf
instability leading to self-pulsing temporal oscillations.

\begin{figure}
\includegraphics[width=8cm]{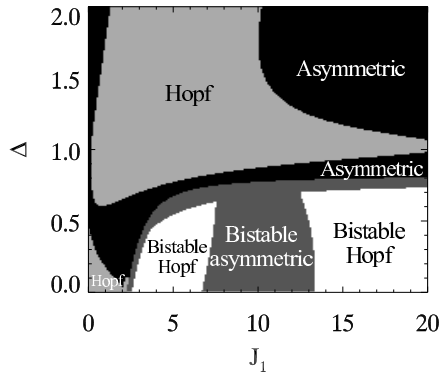}
\caption{Bifurcation diagram showing the stability 
  for $\Delta_1=\Delta_2=\Delta$, $\gamma=0.1$ and $J_2=1.0$. In the
  bistable area the stability of the upper branch is indicated.}
\label{fig:bif-delta-J1}
\end{figure}

The system is well characterized by the relative loss rate $\gamma$,
which in the SHG monomer was shown to determine the degree of
squeezing as well as which field the best squeezing was observed
\cite{collett:1985}. Following the simple layout of the cavity shown
in Fig.~\ref{fig:setup} implies that $\Delta_1=\Delta_2\equiv\Delta$
\cite{detuning}. The bifurcation scenario in the $\{J_1,\Delta\}$
space for $J_2=1.0$ is shown in Fig.~\ref{fig:bif-delta-J1}, which
displays a rich variety of behavior. For $\Delta<1$ bistable behavior
is observed, and the upper branch may be both Hopf unstable as well as
asymmetrically unstable as indicated. For $\Delta>1$ a large Hopf
region is seen, while for $J_1$ large, asymmetric states are observed.

Setting $\gamma=1$ a similar scenario as for $\gamma=0.1$ is observed:
On resonance self-pulsing symmetric states dominates, while bistable
solutions may be seen for $\Delta<-1$ and asymmetric states appear
when $\Delta>0$. For $\gamma=10$ the self-pulsing instability
dominates and asymmetric solutions only appear when detuning is
introduced and simultaneous large values of $J_1$ and $J_2$ are
chosen.  Bistable solutions are not seen here except for very large
coupling strengths, a consequence of the criteria for bistability
(Eq.~(6) in Ref.~\cite{etrich:1997})
\begin{equation}
  \label{eq:bistable}
  \frac{|d_2|(|d_1|-\sqrt{3})}{\sqrt{3}|d_1|+1}>\gamma, \quad d_1d_2>0.
\end{equation}
All these results indicate that the most diverse bifurcation scenario
is when $\gamma\leq 1$.

\section{Photon number spectra}
\label{sec:squeezing-analysis}

The linearized Langevin equation for the positive-$P$ representation can
be used to analytically calculate the spectrum of fluctuations in the
stationary state, provided that the fluctuations are small. These
intracavity fluctuation correlations can be directly related to the
output correlations by using the input-output theory of Gardiner and
Collett \cite{gardiner:1985}. We will only present results in the case
where the symmetric steady states are stable.

The input fields $\hat O_{{\mathrm in},j}(t)$ coupled into the cavity
through the input mirror are posing an instantaneous boundary condition
for the output fields
\begin{equation}
  \label{eq:aout}
  \hat O_{{\mathrm out},j}(t)=\sqrt{2\gamma_j}\hat O_j(t)-\hat
  O_{{\mathrm in},j}(t), \quad \hat O=\hat A, \hat B.
\end{equation}
It should be stressed that in this equation $\gamma_j$ is only the
loss rate of the input mirror, and does not include additional
absorption losses of the cavity that might otherwise have been
included in the Langevin equations. Also note that the input operator
is in the FH taken as a both the classical pump as well as the
fluctuations around this classical level originating from the heat
bath interaction [so really an operator equivalent of the Langevin
input fields from Eqs.~(\ref{eq:noise-W}) and (\ref{eq:noise-P})]. The
fields outside the cavity obey the standard free field commutator
relations
\begin{equation}
  [\hat O_{{\mathrm out},j}(t),\hat O_{{\mathrm out},k}^\dagger(t')] =
  \delta_{jk}\delta(t-t'),
\label{eq:out-comm}
\end{equation}
and
\begin{equation}
  [\hat O_{{\mathrm in},j}(t),\hat O_{{\mathrm in},k}^{\dagger}(t')] =
  \delta_{jk}\delta(t-t'),
\end{equation}
while all other commutators are zero. We want to express correlations
of the out fields entirely on correlations of the intra cavity fields,
hence we want to get rid of terms involving $\hat
O_{{\mathrm in},j}(t)$. Using arguments of causality it may be shown
that this can only be done if time and normally ordered correlations are
considered \cite{gardiner:1985}, as \textit{e.g.}
\begin{mathletters}
  \label{eq:IO-time-normal}
\begin{eqnarray}
  \la \hat A_{{\mathrm out},j}^\dagger(t),\hat
  A_{{\mathrm out},j}(t')\ra = 2 \gamma_j&& \la \hat A_j^\dagger(t),\hat
  A_j(t')\ra 
\\
  \la \hat A_{{\mathrm out},j}(t),\hat
  A_{{\mathrm out},j}(t')\ra = 2 \gamma_j&&\la \hat
  A_j(\mathrm{max}[t,t']),
\nonumber\\
&&\;\;\hat A_j(\mathrm{min}[t,t'])\ra ,
\end{eqnarray}
\end{mathletters}
which precisely implies time and normal order of the correlations.
Since this is exactly what the $P$-representation computes the
intracavity operator averages on the right hand side of
Eq.~(\ref{eq:IO-time-normal}) may be directly replaced by $c$-number
averages from the $P$-representation.

The intensities of the output beams may be found from the photon
number operator $\hat N^O_{{\mathrm out},j}=\hat O_{{\mathrm
    out},j}^\dagger \hat O_{{\mathrm out},j}$. In a photon counting
experiment two-time correlations of the intensities may be calculated as
\end{multicols}
\begin{eqnarray}
  \label{eq:var-tb}
C_{A_jB_k}^{(\pm)}(\tau)&\equiv & \la \hat N^A_{{\mathrm out},j}(t)\pm\hat
  N^B_{{\mathrm out},k}(t), \hat 
  N^A_{{\mathrm out},j}(t+\tau)\pm\hat N^B_{{\mathrm out},k}(t+\tau) \ra 
\nonumber\\
&=& (\la \hat  N^A_{{\mathrm out},j}\ra +\la\hat
  N^B_{{\mathrm out},k}\ra)\delta(\tau)+ 
\la: \hat N^A_{{\mathrm out},j}(t)\pm\hat
  N^B_{{\mathrm out},k}(t), \hat 
  N^A_{{\mathrm out},j}(t+\tau)\pm\hat N^B_{{\mathrm out},k}(t+\tau) :\ra 
\nonumber\\
&=& 2(\gamma_j\la \hat  N^A_{j}\ra +\gamma_k\la\hat N^B_{k}\ra)\delta(\tau)
+ 4\la: \delta \hat N_{A_jB_k}^{(\pm)} (t),\delta \hat N_{A_jB_k}^{(\pm)}
  (t+\tau):\ra , 
\end{eqnarray}
\begin{multicols}{2}
where the notation $\la : \; :\ra$ indicates a normally ordered
average. In the second line we have used the commutator
relations~(\ref{eq:out-comm}) to rewrite to normal order, while the
last line follows from Eq.~(\ref{eq:IO-time-normal}). We have also
introduced 
\begin{equation}
  \delta \hat N_{A_jB_k}^{(\pm)} (t)=\gamma_j\hat N^A_{j}(t)\pm
  \gamma_k\hat N^B_{k}(t) , \quad \hat N^O_j=\hat O_j^\dagger\hat O_j,
\label{eq:deltaN_AB}
\end{equation}
and calculated the variance as $\la S,T\ra\equiv \la ST\ra-\la S\ra
\la T\ra$. 

It is more convenient to investigate these two-time correlations in
the Fourier frequency domain using the Wiener-Khintchine theorem
\cite{mandel:1995} 
\begin{mathletters}
\begin{eqnarray}
    V_{A_jB_k}^{(\pm)}&&(\omega)=\int_{-\infty}^{\infty}d\tau
    e^{i\omega \tau} C_{A_jB_k}^{(\pm)}(\tau)
\nonumber\\
&&=2(\gamma_j\la \hat  N^A_{j}\ra
  +\gamma_k\la\hat   N^B_{k}\ra) 
\nonumber \\
&&+4 \int_{-\infty}^{\infty}d\tau  e^{i\omega \tau}
\la: \delta \hat N_{A_jB_k}^{(\pm)} (t),\delta \hat N_{A_jB_k}^{(\pm)}
  (t+\tau):\ra
\label{eq:V_jk-unscaled}
\\
&&\equiv 
2(\gamma_j\la \hat  N^A_{j}\ra  +\gamma_k\la\hat   N^B_{k}\ra) 
\bar V_{A_jB_k}^{(\pm)}(\omega).
\end{eqnarray}
\end{mathletters}
Here we have introduced the spectrum normalized to shot-noise $ \bar
V_{A_jB_k}^{(\pm)}(\omega)$, and the shot-noise level given by
\begin{equation}
  \label{eq:SN_n}
  C_{SN}=2(\gamma_j\la
\hat N^A_{j}\ra +\gamma_k\la\hat N^B_{k}\ra) ,
\end{equation}
is with this normalization unity. The shot-noise level is the limit
between classical and quantum behavior, hence if $\hat A_j$ and $\hat
B_k$ are coherent states the variance will be $ \bar
V_{A_jB_k}^{(\pm)}(\omega)=1$. A complete violation of the shot-noise
level $ \bar V_{A_jB_k}^{(\pm)}(\omega)=0$ implies that no
fluctuations are associated with the measurement of the intensities
$\hat N^A_{{\mathrm out},j}\pm\hat N^B_{{\mathrm out},k}$. The
correlations between the fields of the same waveguide are
\begin{eqnarray}
  \label{eq:C_tau_mono}
  C_{A_j}(\tau)&\equiv & \la \hat N^A_{{\mathrm out},j}(t), \hat 
  N^A_{{\mathrm out},j}(t+\tau)\ra 
\nonumber\\
&=& 2\gamma_j\la \hat  N^A_{j}\ra \delta(\tau)
+ 4\gamma_j^2\la: \hat N_{A_j} (t), \hat N_{A_j}
  (t+\tau):\ra  ,
\end{eqnarray}
which means that the monomer spectra are
\begin{eqnarray}
      V_{A_j}(\omega)=&&2\gamma_j\la \hat  N^A_{j}\ra
\nonumber\\
&&+4\gamma_j^2 \int_{-\infty}^{\infty}d\tau  e^{i\omega \tau}
\la: \hat N_{A_j} (t),\hat N_{A_j}(t+\tau):\ra,
\label{eq:V_j-unscaled}
\end{eqnarray}
so the shot-noise level is here $C_{SN}=2\gamma_j\la \hat  N^A_{j}\ra$.

Until now everything has been kept in operator form. The next step is
connecting the operator averages with $c$-number classical averages,
which will here be done with the semi-analytical calculations in mind.
Thus, we apply the positive-$P$ representation averages and note that
we shall only consider symmetric states making
${\mathcal A}_j={\mathcal B}_j$, and that the spectra eventually
calculated are linearized.

Expressing the spectra~(\ref{eq:V_jk-unscaled})
and~(\ref{eq:V_j-unscaled}) in dimensionless $c$-numbers from the
$P$-representation we readily have
\begin{mathletters}
\begin{eqnarray}
  V_{A_jB_k}^{(\pm)}&&(\omega)=2n_{s}^{-1}( \bar \gamma_j\bar I_{j} +\bar
  \gamma_k \bar I_{k})   
\nonumber\\
&&+4 \int_{-\infty}^{\infty}d \tau  e^{i \omega  \tau}
\la \delta   I_{A_jB_k}^{(\pm)} (  t),\delta   I_{A_jB_k}^{(\pm)}
  (  t+ \tau)\ra_P
\label{eq:V_jk-scaled}
\\
  V_{A_j}&&(\omega)=2\bar \gamma_jn_{s}^{-1} \bar I_{j}
\nonumber\\
&&+4 \bar \gamma_j^2 \int_{-\infty}^{\infty}d \tau  e^{i \omega  \tau}
\la I_{j}^A (  t), I_{j}^A  (  t+ \tau)\ra_P,
\label{eq:V_j-scaled}
\end{eqnarray}
\end{mathletters}
with the subscript $P$ referring to the averages being calculated in the
$P$-representation. We have here used $\bar\gamma_j=\gamma_j/\gamma_1$
and the $c$-number equivalent of Eq.~(\ref{eq:deltaN_AB})
\begin{equation}
  \label{eq:deltaI_AB}
    \delta I_{A_jB_k}^{(\pm)} (t)=\bar\gamma_jI^A_{j}(t)\pm
  \bar \gamma_kI^B_{k}(t) , \quad I^F_j= F_j^\dagger F_j,
\end{equation}
while the shot-noise level~(\ref{eq:SN_n}) is
\begin{equation}
  \label{eq:SN_p}
  C_{SN}=2n_s^{-1}(\bar \gamma_j \bar I_j+\bar \gamma_k \bar I_k).
\end{equation}
The dimensionless spectra are found by the scalings
\begin{equation}
  \label{eq:V-scale}
\tilde V_{jk}^{(\pm)}(\tilde \omega)=V_{jk}^{(\pm)}(\omega)
\kappa^4/\gamma_1^5, \quad 
\tilde V_{j}(\tilde \omega)=V_{j}(\omega)
\kappa^4/\gamma_1^5  ,
\end{equation}
and tildes have been dropped.

The linearized equations~(\ref{eq:int-pha-lin-mat}) may be solved
directly in frequency space (see Ref.~\cite{reid:1988} for details).
So let us define the spectral matrix of fluctuations in the
$P$-representation in the steady state limit (where we may choose the
time $t$ arbitrarily and henceforth take $t=0$)
\begin{equation}
  \label{eq:S(w)}
  {\mathbf S}^n(\omega)=\int_{-\infty}^{\infty} d\tau e^{i\omega
  \tau}\langle \Delta {\mathbf w} 
  (0) \Delta {\mathbf w}(\tau)^T\rangle_P ,
\end{equation}
with the superscript $n$ indicating that the averages are equivalent
to normally ordered quantum mechanical averages. This may be calculated
using
\begin{equation}
  \label{eq:S-matrix-formula}
  {\mathbf S}^n(\omega)=(-i\omega {\mathbf I}-{\mathbf A})^{-1}{\mathbf D} 
(i\omega {\mathbf I}-{\mathbf A}^T)^{-1},
\end{equation}
where $\mathbf{I}$ is the identity diagonal matrix.

In order to calculate intensity correlations we evaluate terms like
$\la I^A_j(0),I^B_k(\tau)\ra_P$, and expressing this in terms of the
fluctuations around the symmetric steady state, second, third and
fourth order correlations in $\Delta {\mathbf w}$ are obtained. Due to
the strength of the steady state values higher order correlation terms
may be neglected, so we get to leading order
\begin{eqnarray*}
  \la &&I^A_j(0),I^B_k(\tau)\ra_P\simeq {\mathcal A}_j{\mathcal A}_k^*
  \la \Delta A_j^\dagger(0),\Delta B_k(\tau)\ra_P
\nonumber\\
&&+ {\mathcal A}_j^*{\mathcal A}_k \la \Delta
  A_j(0),\Delta B_k^\dagger(\tau)\ra_P
+{\mathcal A}_j^{*}{\mathcal A}_k^*\la \Delta A_j(0),\Delta B_k(\tau)\ra_P
\nonumber\\
&&+{\mathcal A}_j{\mathcal A}_k\la \Delta A_j^\dagger(0),\Delta
  B_k^\dagger(\tau)\ra_P .
\end{eqnarray*}

Using this result the normalized dimensionless
spectra~(\ref{eq:V_jk-scaled}) are
\end{multicols}
\begin{mathletters}
\label{eq:field-dimer-spectra}
\begin{eqnarray}
  \bar V_{A_1A_2}^{(\pm)}(\omega)&=&1+\frac{2}{\bar I_1+\gamma
    \bar I_2}  \Big\{
V_{A_1}^n(\omega)+\gamma^2V_{A_2}^n(\omega)
\pm 2\gamma \Re\big[ 
{\mathcal A}_1^* {\mathcal A}_2 (S^n_{14}(\omega)+S^n_{14}(-\omega)) 
+{\mathcal A}_1^* {\mathcal A}_2^* (S^n_{13}(\omega)+S^n_{31}(\omega))\big]
\Big\}
\\
  \bar V_{A_1B_2}^{(\pm)}(\omega)&=&1+\frac{2}{\bar I_1+\gamma
    \bar I_2}  \Big\{
V_{A_1}^n(\omega)+\gamma^2V_{B_2}^n(\omega)
\pm 2\gamma \Re\big[ 
{\mathcal A}_1^* {\mathcal A}_2 (S^n_{18}(\omega)+S^n_{18}(-\omega)) 
+{\mathcal A}_1^* {\mathcal A}_2^* (S^n_{17}(\omega)+S^n_{71}(\omega))\big]
\Big\}
\\
  \bar V_{A_1B_1}^{(\pm)}(\omega)&=&1+ 
\frac{V_{A_1}^n(\omega)+V_{B_1}^n(\omega)}{\bar I_1}
\pm 2\Re \big[ 
S^n_{16}(\omega)+S^n_{16}(-\omega)
+e^{-i2\phi_1} (S^n_{15}(\omega)+S^n_{51}(\omega))\big]
\\
  \bar V_{A_2B_2}^{(\pm)}(\omega)&=&1+ 
\gamma \frac{V_{A_2}^n(\omega)+V_{B_2}^n(\omega)}{\bar I_2}
\pm 2\gamma\Re\big[ 
S^n_{38}(\omega)+S^n_{38}(-\omega)
+e^{-i2\phi_2} (S^n_{37}(\omega)+S^n_{73}(\omega))\big].
\end{eqnarray}
\end{mathletters}
\begin{multicols}{2}
Here we have used the normally ordered (indicated with a superscript
$n$) single mode spectrum  defined as  
\begin{equation}
  \label{eq:V_N}
    V_{A_j}^n(\omega)= \int_{-\infty}^{\infty}d \tau  e^{i \omega  \tau}
\la I_{j}^A (0), I_{j}^A  ( \tau)\ra_P,
\end{equation}
so
\begin{mathletters}
\label{eq:field-spectra-no}
\begin{eqnarray}
  V_{A_1}^n(\omega)&=&\bar I_1\Big( S^n_{12}(\omega)+S^n_{12}(-\omega)
\nonumber\\&&
\quad +2\Re[S^n_{11}(\omega)e^{-i2\phi_1}]
\Big)
\\
  V_{A_2}^n(\omega)&=&\bar I_2\Big( S^n_{34}(\omega)+S^n_{34}(-\omega)
\nonumber\\&&
\quad +2\Re[S^n_{33}(\omega)e^{-i2\phi_2}]
\Big),
\end{eqnarray}
\end{mathletters}
and $V_{B_j}^n(\omega)=V_{A_j}^n(\omega)$. With these quantities the
monomer spectra~(\ref{eq:V_j-scaled}) normalized to shot-noise are
readily calculated
\begin{eqnarray}
\label{eq:field-spectra-monomer}
\bar V_{A_j}(\omega)&=&1+\frac{2\bar\gamma_j}{\bar
  I_j}V_{A_j}^n(\omega) .
\end{eqnarray}
The calculations of the spectra use the general symmetry properties of
the spectral matrix ${\mathbf S}^n(\omega)$, so \textit{e.g.}
$S^n_{11}(\omega)=[S^n_{22}(\omega)]^*$ and
$S^n_{12}(\omega)=S^n_{21}(-\omega)$.

\section{Results}
\label{sec:Results}

In this section we present intensity correlation spectra both from the
semi-analytical derivation, as well as results from the numerical
simulations (the numerical method is discussed in
App.~\ref{sec:num}). The chosen examples are only illustrative for the
overall behavior, and the results hold for large parameter
areas. This is especially important to stress for the coupling
parameters, since they are not so easy to control experimentally as the
detunings and loss rates.

In order to understand the effect of the coupling between the
waveguides, a comparison to the results of the single waveguide will
be made. It is important to distinguish between two cases: a) The
monomer correlations, where we talk about the correlations between the
fields within a single waveguide given by
Eq.~(\ref{eq:field-spectra-monomer}) and where coupling is still
present. b) The limit of no coupling, where the spectra will behave as
a single isolated waveguide. This limit is important since it allows
us to compare with the results previously obtained by Collett and
Walls \cite{collett:1985}, and henceforth this limit is referred to as
the SHG monomer. Finally, we denote the spectral variances $\bar
V_{A_jB_k}^{(\pm)}(\omega)$ as the dimer correlations or variances.

It was shown by Collett and Walls \cite{collett:1985} that in the SHG
monomer without detuning very good squeezing in the fundamental
quadrature $-i(\hat A_1 e^{-i\theta_1}-\hat A_1^\dagger
e^{i\theta_1})$ is obtained when $\gamma$ is small, and conversely
when $\gamma$ is large good squeezing in the second-harmonic
quadrature $-i(\hat A_2 e^{-i\theta_2}-\hat A_2^\dagger
e^{i\theta_2})$ is observed. These squeezing spectra were optimized by
choosing a proper value of the quadrature phase and as it turns out
$\theta_1=\theta_2=\pi/2$ maximizes the squeezing in both cases
(corresponding to the amplitude quadrature). For exactly this value
the quadrature correlations coincide (to leading order) with the
monomer intensity correlations~(\ref{eq:C_tau_mono}) so the results of
Ref.~\cite{collett:1985} also predicts excellent noise suppression in
the monomer photon number variances considered in this paper. Note
that the choice $\theta=\pi/2$ only maximizes the squeezing when
detuning is zero, as it was shown by Olsen \textit{et al.}
\cite{olsen:1999}.

Generally, the violation of the classical (or shot-noise) limit
requires that the fluctuations diverge in a given observable of the
fields. When this happens the spectral variance for this observable
becomes large, and the canonical conjugate observable of the fields
will in turn have a small variance as a consequence of the Heisenberg
uncertainty relation of canonical conjugate observables. A typical
situation where the fluctuations diverge is close to a transition from
one stable state to another, and therefore violations of the
shot-noise limit is normally studied close to bifurcation points. In
this paper we study the sum and the difference of the intensities of
the fields, so a violation of the classical limit implies that
sub-Poissonian statistics is observed and that the photons at the
photodetectors are anti-bunched; they arrive more regularly than if
coherent beam intensities (which obey Poissonian statistics) were
measured. The problem with the intensity observable is to find the
conjugate observable in which the fluctuations should become large
when the intensity correlations violate the classical limit. Numerous
attempts to create the most intuitive conjugate observable, namely a
phase operator, has not been entirely successful
\cite{mandel:1995}. On the other hand, in a photon-counting experiment
it is exactly intensity correlations that are measured, making them a
suitable choice for a direct experimental implementation.

The analytical and numerical results presented in the following
display excellent mutual agreement. In order to achieve this it was
necessary to have the time resolution of the two-time correlations low
enough to describe the temporal variations, while simultaneously
keeping upper limit of the correlation time (corresponding to the
limits $\tau\rightarrow\pm\infty$ of the analytical integral) long
enough for the two-time correlation to become close to zero.
Otherwise, the temporal Fourier transform of the correlations will
give spectra that are in disagreement with the analytical results.
Needless to say, this had to be checked for each case as the
parameters were varied, but generally we used $N=512$ or 1024 points
with a resolution in the range $\Delta \tau=0.04-0.2$ to calculate the
two-time correlations $C(\tau)$. Finally, the length of the
simulations was around $10^6$ time units before the correlations
converged to the degree shown in the following.

\begin{figure}
\includegraphics[width=8cm]{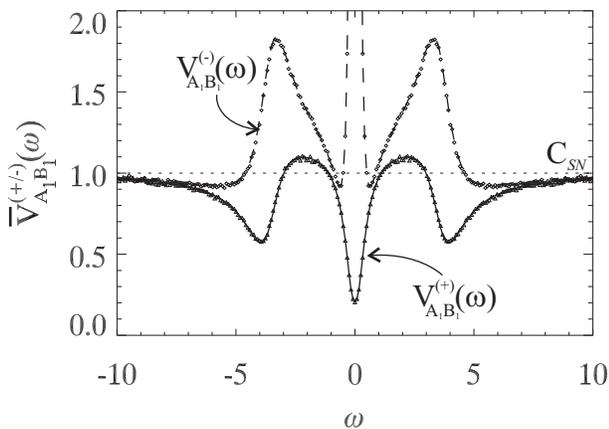}
\caption{Photon number spectra $\bar V_{A_1B_1}^{(\pm)}(\omega)$ for the
  parameters in Fig.~\ref{fig:bif-delta-J1} and $J_1=3.0$, $\Delta=0$ and
  $E=3.275$, on the lower branch just before a bistable turning point.
  Lines show analytical results while points are numerical results. The
  shot-noise level is indicated with $C_{SN}$.}
\label{fig:0.1_bistable_tp}
\end{figure}

\subsection{$\gamma$ small}
\label{sec:gamma-small}

First we consider the case where $\gamma$ is small and good
nonclassical correlations in the FH fields are expected ($\bar
V_{A_1}(\omega)\simeq\gamma/[1+\gamma]$ close to self-pulsing
transitions) \cite{collett:1985}.  For $\gamma=0.1$ we observed the
strongest violations of the quantum limit close to bistable turning
points. An example is shown in Fig.~\ref{fig:0.1_bistable_tp} located
in the bistable region of Fig.~\ref{fig:bif-delta-J1}, and the system
is set on the lower branch of the bistable curve just before the right
turning point. The dimer spectrum of the sum of the FH fields shows a
near Lorentzian dip in the region of $\omega=0$ that goes down to
$\bar V_{A_1B_1}^{(+)}(\omega)\simeq 0.2$, implying strong twin-beam
correlations, while the FH difference shows excess noise here. Taking
$\gamma$ even smaller we were able to get $\bar
V_{A_1B_1}^{(+)}(\omega)$ very close to zero in the presence of
bistable turning points, a behavior similar to the
$\gamma/(1+\gamma)$ behavior observed in the SHG monomer close to
self-pulsing transitions. The excellent correlations are only seen
close to the bistable transition, taking \textit{e.g.}  $E=3.2$ for
the parameters in Fig.~\ref{fig:0.1_bistable_tp}, the minimum of the
spectrum is $\bar V_{A_1B_1}^{(+)}(\omega)\simeq 0.35$.  Returning to
Fig.~\ref{fig:0.1_bistable_tp}, at $\omega\simeq 4$ the FH sum
spectrum again shows nonclassical correlations of approximately 60\%
of the shot-noise limit. The frequency almost coincides with the
frequency of the Hopf instability that is suppressed here; the
eigenvalue with the largest real part is the bistable one while the
eigenvalues with more negative real parts have $\Im(\lambda)\simeq
3.6$.

Here it is relevant to mention that bistability is also present in the
SHG monomer \cite{etrich:1997,etrich:1997b} (as
Eq.~(\ref{eq:bistable}) indicates this requires nonzero detunings with
equal sign), and to the best of our knowledge nobody has here
investigated the quantum behavior. Let us write the detunings of the
SHG monomer as $\bar \Delta_j$. Due to the invariance of the symmetric
steady state solutions when $\bar \Delta_j=\Delta_j-J_j$ we can obtain
the same bistable state investigated in Fig.~\ref{fig:0.1_bistable_tp}
in the SHG monomer by setting $\bar \Delta_1=-3.0$ and $\bar
\Delta_2=-1.0$. The spectrum $\bar V_{A_1}(\omega)$ displays here
exactly the same behavior as $\bar V_{A_1B_1}^{(+)}(\omega)$ in
Fig.~\ref{fig:0.1_bistable_tp}, so also in the SHG monomer perfect
anti-bunching behavior may be obtained in the small $\gamma$ limit.
Generally, the dimer spectra $\bar V_{A_iB_i}^{(+)}(\omega)$ can be
reproduced by the monomer spectra $\bar V_{A_i}(\omega)$ when taking
$\bar \Delta_j=\Delta_j-J_j$. This is not valid, however, close to a
transition to asymmetric states, and also the spectra $\bar
V_{A_iB_i}^{(-)}(\omega)$ have no equivalents in the no-coupling
intensity correlations.

\begin{figure}
\includegraphics[width=8cm]{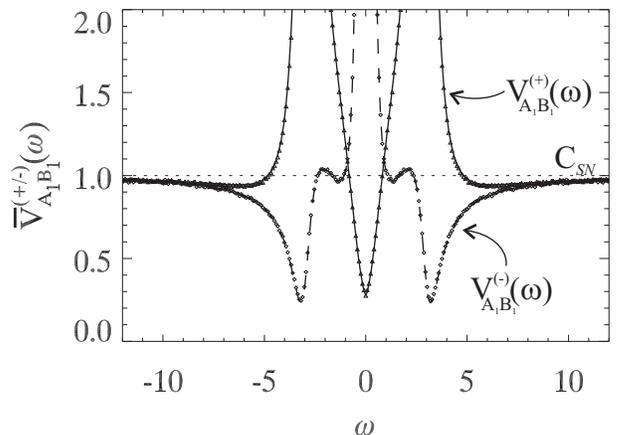}
\caption{Photon number spectra $\bar V_{A_1B_1}^{(\pm)}(\omega)$ for the
  parameters in Fig.~\ref{fig:bif-delta-J1}, $J_1=2.0$, $\Delta=0$
  and $E=2.3$. Lines show analytical results while points are
  numerical results. The shot-noise level is indicated with $C_{SN}$.}
\label{fig:0.1_bistable}
\end{figure}

In the self-pulsing region of Fig.~\ref{fig:bif-delta-J1} it is
possible to obtain good correlations if the system is set close to the
bistable area. The spectra in Fig.~\ref{fig:0.1_bistable} are for a
pump value where both the bistable and the self-pulsing eigenvalues
are of approximately the same strength, and the plot shows that the
dimer spectrum of the FH difference have strong noise suppression for
nonzero $\omega$, that originates from the self-pulsing instability
setting in at $E_{\mathrm SP}=4.7$.  Also the sum shows strong noise
suppression now at $\omega=0$, caused by the proximity of the bistable
area which gives rise to an eigenvalue with $\Im(\lambda)=0$ that
never has $\Re(\lambda)>0$. The good correlations observed here are
apparently a result of a competition between the bistable state and
the emerging self-pulsing instability, that eventually dominates for
higher pump levels. When the self-pulsing threshold is approached, the
nonclassical behavior is less pronounced. This does not necessarily
mean that nonclassical states are not present here, but probably that
the intensity is the wrong observable in which to observe nonclassical
correlations here.  Such a case was observed in the SHG monomer, where
it was found that detuning of the system makes the squeezing ellipse
turn so the best squeezing is no longer observed in the amplitude
quadrature \cite{olsen:1999}.

\begin{figure}
\includegraphics[width=8cm]{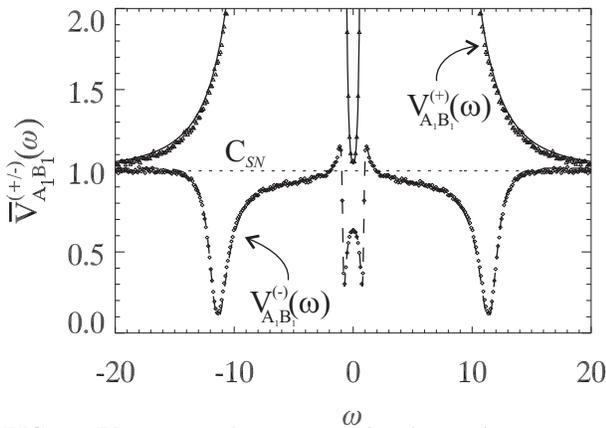}
\caption{Photon number spectra for $\Delta_1=\Delta_2=1.1$ and
  $\gamma=0.1$, $J_1=20.0$, $J_2=1$ and $E/E_{\mathrm sym}=0.97$.
  Lines show analytical results while points are numerical results.
  The shot-noise level is indicated with $C_{SN}$.}
\label{fig:0.1_sym}
\end{figure}

When detuning is introduced, the system can become asymmetrically
unstable for certain couplings as was shown in
Fig.~\ref{fig:bif-delta-J1}.  Generally, the asymmetric instability
shows sub-Poissonian twin-beam correlations in the dimer FH difference
spectrum, which is especially pronounced when $J_1\gg J_2$ where
almost perfect anti-bunching was observed. In Fig.~\ref{fig:0.1_sym}
the dimer spectra $\bar V_{A_1B_1}^{(\pm)}(\omega)$ are shown for
$\Delta=1.1$, $J_1=20$ and $J_2=1$, taken close to the symmetric
transition $E_{\mathrm sym}$, and intensity correlations until 8\% of
the shot-noise limit is seen in the FH difference correlations at
nonzero $\omega$. By carefully selecting the parameters we were even
able to see correlations until 3\% of the shot-noise level, which
underlines that excellent nonclassical correlations are observed here.
This result is quite robust; good sub-Poissonian correlations are
observed also further below the transition as well as for considerably
lower values of the FH coupling strength, while the peaked structure
around $\omega=0$ is quite sensitive to the pump level since it is not
seen taking the system even closer to $E_{\mathrm sym}$. We note from
Fig.~\ref{fig:bif-delta-J1} the presence of the self-pulsing
instability for the parameters chosen, and even if quite good
correlations are observed inside the large asymmetric area for $J_1$
large, the best results are obtained close to the self-pulsing
regions. Again this shows that two competing instabilities appear to
give rise to strong nonclassical correlations.  We note finally that
the frequency $\omega\simeq 11$, where the best correlations in
Fig.~\ref{fig:0.1_sym} are observed, almost coincides with the
frequency of the self-pulsing eigenvalue that is damped the most.
Thus, paradoxically, in this case the least dominating eigenvalue is
determining the frequency of the strong correlations.

\begin{figure}
\includegraphics[width=8cm]{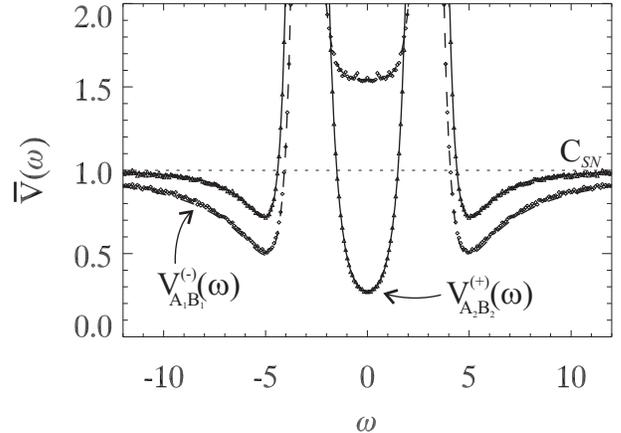}
\caption{Photon number spectra on resonance and for $\gamma=1$ and
  $J_1=4.0$, $J_2=1.0$ and $E/E_{\mathrm SP}=0.95$.  Lines show
  analytical results while points are numerical results.  The
  shot-noise level is indicated with $C_{SN}$.}
\label{fig:1_hopf}
\end{figure}

\subsection{$\gamma=1$}
\label{sec:gamma=1}

Setting the loss rates to be identical, $\gamma=1$, the self-pulsing
instability gives rise to strong nonclassical correlations all the way
to the transition to the self-pulsing state, while the bistable
transition displays only weak violations. As an example of the
self-pulsing correlations Fig.~\ref{fig:1_hopf} displays selected
spectra for the dimer correlations. The sum of the SH fields displays
strong correlations at $\omega=0$, which goes to 25\% of the
shot-noise limit when $E\rightarrow E_{\mathrm SP}$, a result that can
also be found in the SHG monomer by introducing equivalent detunings.
Also the dimer FH difference correlations are strong; for nonzero
$\omega$ correlations around 50\% of the shot-noise limit are seen.

\subsection{$\gamma$ large}
\label{sec:gamma-large}

\begin{figure}
\includegraphics[width=8cm]{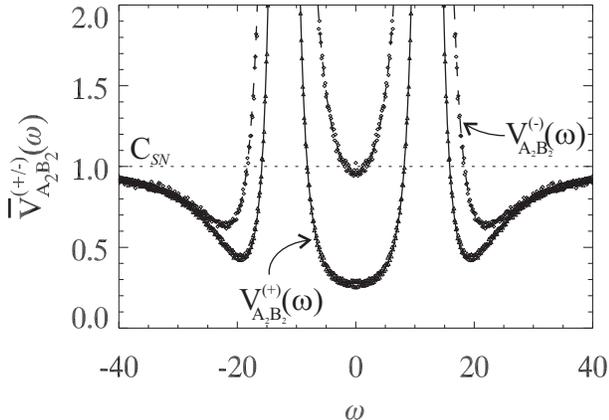}
\caption{Photon number spectra on resonance and for $\gamma=10$ and
  $J_1=6.0$, $J_2=2.0$, and $E/E_{\mathrm SP}=0.95$.  Lines show
  analytical results while points are numerical results.  The
  shot-noise level is indicated with $C_{SN}$.}
\label{fig:10_hopf}
\end{figure}

For large $\gamma$ the SHG monomer predicts strong shot-noise
violations in the SH spectrum at the self-pulsing threshold
\cite{collett:1985} ($\bar V_{A_2}(\omega)\simeq(1+\gamma)^{-1}$ close
to self-pulsing transitions), and for the dimer similar levels of
correlations can be obtained close to the self-pulsing transition. As
an example of the behavior Fig.~\ref{fig:10_hopf} shows highly
nonclassical twin-beam correlations in the dimer spectrum for the sum
of the SH intensities. For the selected parameters we observe also
sub-Poissonian behavior in $\bar V_{A_2B_2}^{(-)}(\omega)$, which, in
contrast to $\bar V_{A_2B_2}^{(+)}(\omega)$, cannot be observed in the
SHG monomer with equivalent detunings.  Although asymmetric areas were
found for nonzero detunings, only weak sub-Poissonian correlations
were observed there in the difference of the SH fields (the sum
correlations still show strong sub-Poissonian correlations here, as
expected from the SHG monomer predictions).

\section{Summary and discussion}
\label{sec:summary}

In this paper we have proposed a model we denote the \textit{quantum
  optical dimer} for studying the effects of a simple mode coupling in a
cavity. The model consisted of two $\chi^{(2)}$ nonlinear waveguides in
a cavity, with coupling between them from evanescent overlapping
waves that was assumed weak and linear. We chose to restrict ourselves
to investigating nonclassical correlations, and hence derived the
nonlinear quantum equations of evolution for the system, resulting in
a set of stochastic Langevin equations.

Using a linearized analysis we showed that the system for low pump
levels allowed a symmetric state to be stable, where both waveguides
are in the same stable state. Depending on the system parameters this
state destabilized into a self-pulsing state, where temporal
oscillations are observed, or bistable solutions in the steady states
occurred. For some parameters, the symmetric steady state lost
stability in favor of an asymmetric steady state, in which the
two waveguides has different steady state solutions. 

We investigated the effects of the quantum noise present in the system
by calculating two-time intensity correlation spectra of the output
fields. It was shown that sub-Poissonian correlations were present in
the system, especially when considering the sum or difference of the
field intensities from each waveguide implying strong twin-beam
correlations. This nonclassical anti-bunching effect is a true
manifestation of a quantum behavior and was observed mainly in three
cases:
\begin{itemize}
\item Close to bistable turning points the strongest violations of the
  classical limit were observed in the spectrum at $\omega=0$,
  corresponding to correlations at infinite time. In the limit of
  $\gamma\ll 1$ perfect noise suppression could be obtained in the sum
  of the FH intensities, resulting in perfect twin-beam
  behavior. This was also observed in the single waveguide model,
  when equivalent detunings were introduced.
\item Close to threshold for self-pulsing behavior strong twin-beam
  correlations were observed both at $\omega=0$ and also at values of
  $\omega$ coinciding with the oscillation frequency of the emerging
  instability. Thus, the correlations here anticipate the behavior
  above the threshold analogously to the idea of a quantum image
  \cite{lugiato:1995}, where the spatial modulations are encoded in
  the correlations below threshold while the average intensity remains
  homogeneous.
\item Excellent sub-Poissonian correlations were observed close to
  transitions from symmetric to asymmetric steady states, which is a
  wholly unique transition to the dimer. This instability is closely
  related to the near-field of a modulationally unstable system in
  presence of diffraction; the dimer sites could be thought of as
  neighboring near-field pixels. Variances down to 3\% of the
  shot-noise level were observed in the the difference of the FH
  fields, implying nearly perfect twin-beam behavior. The
  correlations were particularly strong when the FH coupling strength
  was much larger than the SH and when the FH loss rate was much
  larger than the SH loss rate. These results indicate that strong
  near-field correlations can be observed in SHG with diffraction
  close to a modulational instability, which to our knowledge has not
  been investigated yet.
\end{itemize}
It is worth noting that the twin-beam correlations reported here were
all originating from the dimer coupling across the waveguides; while
each field was created individually from the nonlinear interaction in
the corresponding waveguide, the coupling between the waveguides gave
rise to the strong nonclassical twin-beam correlations. Hence, the
nonclassical correlations arise not because the photons are twins, but
rather because they are ``brothers''.

Common for all these cases was that distinctively strong nonclassical
correlations were observed in parameter regimes where two types of
instabilities were competing. This was observed in self-pulsing areas
close to bistable and asymmetric regimes, and also in asymmetric areas
close to bistable and self-pulsing regimes. These enhanced
correlations from competing instabilities has to the best of our
knowledge not been reported before.

We showed that the relative input mirror loss rate
$\gamma=\gamma_2/\gamma_1$ between the SH and FH fields had a strong
influence on the sub-Poissonian behavior, as was previously shown by
Collett and Walls \cite{collett:1985} in the SHG monomer. For small
$\gamma$ the strongest nonclassical states were mainly observed in the
FH fields while for large $\gamma$ they were mainly observed in the SH
fields.  Since the photon life times in the cavity are inversely
proportional to the loss rates of the input mirror, the time spent in
the cavity is decisive for the level of nonclassical correlations of
the output fields; the field with the shortest time spent in the
cavity displays the strongest nonclassical correlations.  The fact
that a long interaction time tends to destroy the nonclassical
correlations have also been observed in propagation setups,
specifically Olsen \textit{et al.} \cite{olsen:2000} showed that in
propagation SHG the presence of quantum fluctuations caused a dramatic
revival of the FH after a certain propagation length, causing the
variance to go above shot-noise level.

We stress that the results presented here were very robust to changes
in the parameters. Thus, large parameters areas exist where strong
nonclassical behavior can be seen. This is especially important to
stress for the coupling parameters, since they are not so easily
controlled experimentally.

We investigated only the symmetric state of the system, and a future
study of the stability, dynamics and nonclassical properties of the
asymmetric states is relevant. In this context we should also mention
that the coupling between the waveguides chosen here was conservative
(imaginary coupling in the Langevin equations) while it could also
have been dissipative (real) or a combination (complex). This highly
depends on the actual setup, and a future study should include the
possibility of a general complex coupling.

\section{Acknowledgements}
\label{sec:ack}

Financial support from Danish National Science Research Council grant
no. 9901384 is gratefully acknowledged. We thank Pierre Scotto,
Roberta Zambrini, Peter Lodahl, Jens Juul-Rasmussen, Mark Saffman and
Ole Bang for fruitful discussions. Y.G. thanks MIDIT and Informatics
and Mathematical Modelling at the Technical University of Denmark for
financial support.

\appendix

\section{The quantum-to-classical correspondence}
\label{sec:Fokk-Planck-descr}

In this section we show how the master equation~(\ref{eq:ME}) is
converted into an equivalent partial differential equation by
expanding the density matrix in coherent states weighted by a
quasi-probability distribution \cite{walls:1994,carmichael}. In
this distribution the operators are replaced by equivalent $c$-numbers,
where the particular correspondence between these depends on the
ordering of the operators. In the case where only up to second order
derivatives appear in the corresponding partial differential equation,
the equation is on Fokker-Planck form allowing equivalent sets
of stochastic Langevin equations to be found.

We are now left with a choice of probability distribution, be it
either the $P$, Wigner or $Q$ distribution giving normal, symmetric or
anti-normal averages, respectively. The aim of this paper is to
calculate two-time correlation spectra outside the cavity, and in
order to do this most conveniently, the moments of the intracavity
fields must be time and normally ordered, cf. the discussion in
Sec.~\ref{sec:squeezing-analysis}. Since the $P$-representation will
immediately give the time and normally ordered averages needed, this is
the favorable representation in this context. As will be explained
later we use the Wigner representation for the numerical simulations,
and therefore we now provide a general way of deriving the QPD equations.

The approach we use is to introduce a characteristic function 
\begin{equation}
  \label{eq:char-func}
  \chi(z)=\Tr[\hat D(z)\hat \rho ],
\end{equation}
so $\chi$ is the trace over a displacement operator $\hat D(z)$ acting
on the density matrix [i.e. the expectation value of $\hat
D(z)$]. The choice of ordering now amounts to choosing the ordering of
$\hat D(z)$. In the symmetric ordering
\begin{equation}
  \label{eq:D-symm}
  \hat D_s(z)=e^{z \hat A^\dagger - z^* \hat A},
\end{equation}
where $z$ is a complex number describing the amplitude of a coherent
field, and $\hat A$ is a boson operator. The normally ordered
displacement operator is
\begin{equation}
  \label{eq:D-norm}
  \hat D_n(z)=e^{z \hat A^\dagger}e^{ - z^* \hat A},
\end{equation}
and the anti-normally ordered displacement operator is
\begin{equation}
  \label{eq:D-antinorm}
  \hat D_a(z)=e^{-z^* \hat A}e^{z \hat A^\dagger}.
\end{equation}
The QPD is now given as a Fourier transform of the characteristic
function
\begin{equation}
  \label{eq:W(alpha)}
  W(\alpha)=\int d^2z\chi(z)e^{z^*\alpha - z\alpha^*},
\end{equation}
where the integration measure $d^2z$ means integration over the entire
complex plane. From this relation an equivalence between operators
and $c$-numbers has been established as $\hat A \leftrightarrow \alpha$
and $\hat A^\dagger \leftrightarrow \alpha^*$. The $c$-number averages
may now be calculated as, \textit{e.g.},
\begin{equation*}
  \la \alpha^{*} \alpha\ra=\int d^2\alpha W(\alpha)\alpha^*\alpha,
\end{equation*}
and obviously since the operator ordering determines the specific
form of $W(\alpha)$ from Eq.~(\ref{eq:W(alpha)}), this in turn implies
that the $c$-number averages also are influenced by the choice of
ordering.

From differentiating Eq.~(\ref{eq:char-func}) with respect to time we
get
\begin{equation}
  \frac{\partial \chi(z)}{\partial t}=\Tr\left(\hat D(z)\frac{\partial
  \hat \rho}{\partial t}\right),
\label{eq:W-chi-dt}
\end{equation}
with $\partial_t\hat\rho$ being governed by the master
equation~(\ref{eq:ME}), and we have used that in the Schr\"odinger
picture the operators [here $\hat D(z)$] are independent of time. The
approach is now to differentiate $\hat D(z)$ with respect to
\textit{e.g.} $z$ and rearrange to get
\begin{equation}
  \label{eq:a+D}
  \hat A^\dagger \hat D_s(z)=\left(\frac{\partial}{\partial
  z}+\frac{z^*}{2}\right)\hat D_s (z).
\end{equation}
The right hand side of Eq.~(\ref{eq:W-chi-dt}) is evaluated using
Eq.~(\ref{eq:a+D}) and the similar other expressions.
Eq.~(\ref{eq:W-chi-dt}) is then Fourier transformed according to
Eq.~(\ref{eq:W(alpha)}), assuming the characteristic function is well
behaved, to finally give the equation governing the time evolution of
$W(\alpha)$.

Choosing the normally ordered displacement operator given by
Eq.~(\ref{eq:D-norm}) the equation for the Glauber-Sudarshan
$P$-representation is derived, which is on Fokker-Planck form.
However, due to problems with negative diffusion in quantum optics the
generalized $P$-distributions \cite{drummond:1980a,yuen:1986} are
normally used instead, where the problems are surpassed by doubling
the phase space. We will use the positive $P$-representation, which
can be derived by replacing all $\alpha_j^*\rightarrow
\alpha_j^\dagger$ and $\beta_j^*\rightarrow \beta_j^\dagger$ in the
Fokker-Planck equation of the Glauber-Sudarshan $P$-representation.
This means that $\alpha_j^\dagger$ is now an independent complex
quantity instead of being the complex conjugate of $\alpha_j$. The
Fokker-Planck equation using the positive $P$-representation
corresponding to the master equation~(\ref{eq:ME}) is then
\begin{eqnarray}
  \frac{\partial W_n({\mathbf x})}{\partial t}&=&\Bigg\{
  \frac{\partial}{\partial \alpha_1}\left[\alpha_1(\gamma_1 -
    i\delta_1) +i J_1 \beta_1-\kappa
  \alpha_1^\dagger\alpha_2-\mathcal{E}_{\mathrm{p},a} 
  \right] 
  \nonumber\\  &+&
\frac{\partial}{\partial \alpha_1^\dagger} \left[\alpha_1^\dagger
  (\gamma_1 + i\delta_1) -iJ_1\beta_1^\dagger-\kappa
  \alpha_1\alpha_2^\dagger-\mathcal{E}_{\mathrm{p},a}^*\right] 
  \nonumber\\
  &+&\frac{\partial}{\partial \alpha_2}\left[\alpha_2(\gamma_2 -
    i\delta_2) +i J_2 \beta_2+\frac{\kappa}{2} {\alpha_1}^2 \right] 
  \nonumber\\  &+&
\frac{\partial}{\partial \alpha_2^\dagger} \left[\alpha_2^\dagger
  (\gamma_2 + i\delta_2) -iJ_2\beta_2^\dagger+\frac{\kappa}{2}
  {\alpha_1^\dagger}^2 \right] 
  \nonumber\\
  &+&\frac{\partial}{\partial \beta_1}\left[\beta_1(\gamma_1 -
    i\delta_1) +iJ_1\alpha_1 -\kappa \beta_1^\dagger \beta_2
  -\mathcal{E}_{\mathrm{p},b}\right] 
  \nonumber\\  &+&
\frac{\partial}{\partial \beta_1^\dagger} \left[\beta_1^\dagger
  (\gamma_1 + i\delta_1) -iJ_1\alpha_1^\dagger -\kappa
  \beta_1\beta_2^\dagger-\mathcal{E}_{\mathrm{p},b}^*\right] 
  \nonumber\\
  &+&\frac{\partial}{\partial \beta_2}\left[\beta_2(\gamma_2 -
    i\delta_2) +iJ_2\alpha_2 +\frac{\kappa}{2} \beta_1^2 \right] 
  \nonumber\\  &+&
\frac{\partial}{\partial \beta_2^\dagger} \left[\beta_2^\dagger
  (\gamma_2 + i\delta_2) -iJ_2\alpha_2^\dagger -\frac{\kappa}{2}
  {\beta_2^\dagger}^2 \right]  
  \nonumber\\
  &+&\frac{\kappa}{2}\Big[\frac{\partial^2}{\partial
      \alpha_1^2}\alpha_2 +\frac{\partial^2}{\partial
      {\alpha_1^\dagger}^2}\alpha_2^\dagger
  \nonumber\\
  &&
    \;\;\;+\frac{\partial^2}{\partial
      \beta_1^2}\beta_2 +\frac{\partial^2}{\partial
      {\beta_1^\dagger}^2}\beta_2^\dagger \Big] 
\Bigg\}W_n({\mathbf x}),
\label{eq:FPE-chi2-dimer}
\end{eqnarray}
where the $c$-number equivalents of the operators are $\{\hat A_j,\hat
A_j^\dagger\} \leftrightarrow \{\alpha_j,\alpha_j^\dagger\}$ and
$\{\hat B_j,\hat B_j^\dagger\} \leftrightarrow
\{\beta_j,\beta_j^\dagger\}$, and ${\mathbf x}$ represents the
$c$-number states
\begin{eqnarray}
{\mathbf x}=\{\alpha_1,\alpha_1^\dagger,\alpha_2,\alpha_2^\dagger,
\beta_1,\beta_1^\dagger,\beta_2,\beta_2^\dagger\}. 
\nonumber
\end{eqnarray}

The numerical simulation of the positive $P$-representation has been
reported as very difficult, mainly due to divergent trajectories
\cite{gilchrist:1997}, cf. the discussion in App.~\ref{sec:num}.
Therefore we choose to use the Wigner representation for the numerical
simulations, obtained by using the symmetric displacement
operator~(\ref{eq:D-symm}). The time evolution of the Wigner
distribution is governed by
\begin{eqnarray}
\label{eq:c-number-Wigner-dimer}
  \frac{\partial W_s({\mathbf x})}{\partial t}&=&
\Bigg\{
\frac{\partial}{\partial \alpha_1}[(\gamma_1-i\delta_1)\alpha_1
  -\kappa\alpha_1^*\alpha_2+iJ_1\beta_1-\mathcal{E}_{\mathrm{p},a}] 
\nonumber\\&+&
\frac{\partial}{\partial \alpha_2}[(\gamma_2-i\delta_2)\alpha_2
  +\frac{\kappa}{2}\alpha_1^2 + iJ_2\beta_2] 
\nonumber\\
&+&\frac{\partial}{\partial \beta_1}[(\gamma_1-i\delta_1)\beta_1
  -\kappa\beta_1^*\beta_2+iJ_1\alpha_1-\mathcal{E}_{\mathrm{p},b}] 
\nonumber\\
&+&\frac{\partial}{\partial \beta_2}[(\gamma_2-i\delta_2)\beta_2
  +\frac{\kappa}{2}\beta_1^2+iJ_2\alpha_2] 
\nonumber\\
&+&\frac{\gamma_1}{2}
\left(\frac{\partial^2}{\partial\alpha_1 \partial 
    \alpha_1^*} + \frac{\partial^2}{\partial\beta_1 \partial
    \beta_1^*}\right)
\nonumber\\
&+&\frac{\gamma_2}{2}
\left(\frac{\partial^2}{\partial\alpha_2 \partial 
    \alpha_2^*} + \frac{\partial^2}{\partial\beta_2 \partial
    \beta_2^*}\right)
\nonumber\\
&+&\frac{\kappa}{4}\left(
  \frac{\partial^3}{\partial \alpha_1^2 \partial\alpha_2^*}  
  +\frac{\partial^3}{\partial \beta_1^2 \partial \beta_2^*}  
\right) +c.c.  
\Bigg\}W_s({\mathbf x}),
\end{eqnarray}
where the $c$-number equivalents of the operators are $\{\hat A_j,\hat
A_j^\dagger\} \leftrightarrow \{\alpha_j,\alpha_j^*\}$ and $\{\hat
B_j,\hat B_j^\dagger\} \leftrightarrow \{\beta_j,\beta_j^*\}$ and
\begin{eqnarray}
{\mathbf x}=\{\alpha_1,\alpha_1^*,\alpha_2,\alpha_2^*,
\beta_1,\beta_1^*,\beta_2,\beta_2^*\} .
\nonumber
\end{eqnarray}
Due to the third order derivatives
Eq.~(\ref{eq:c-number-Wigner-dimer}) is not on Fokker-Planck form, a
problem we address in App.~\ref{sec:num}.  Note that the $+c.c.$ term
(denoting the complex conjugate) at the end applies to the entire
equation.

The connection from the QPD equations to the  stochastic Langevin
equations can be made if the QPD equation is on Fokker-Planck
form, which for a system with $m$ $c$-number states $x_j$ is
\cite{carmichael} 
\begin{eqnarray}
  \label{eq:FPE-general1}
  \frac{\partial W({\mathbf x})}{\partial t}= \Big\{&&
    -\sum_{j=1}^m \frac{\partial}{\partial x_j} A_j({\mathbf x})
\nonumber\\&&    
+\frac{1}{2}\sum_{j,k=1}^m \frac{\partial^2}{\partial x_j \partial
      x_k}
    D_{jk}({\mathbf x})\Big\}W({\mathbf x}),
\end{eqnarray}
Using Ito rules for stochastic integration the equivalent set of
Langevin equations is
\begin{equation}
  \label{eq:lang-ito-general-simple}
  \frac{\partial x_j}{\partial t}=A_j({\mathbf x})
  +w_j(t),
\end{equation}
where $w_j(t)$ are Gaussian white noise terms, delta correlated in time
according to the diffusion matrix ${\mathbf D}$
\begin{equation}
  \label{eq:Gaussian-noise-ito}
  \langle w_j(t)w_k(t')\rangle=D_{jk}({\mathbf x})\delta(t-t').
\end{equation}
We note that if ${\mathbf D}$ depends on ${\mathbf x}$ the noise is
labeled multiplicative which is the case for the positive-$P$
Eq.~(\ref{eq:FPE-chi2-dimer}), otherwise it is additive as it is for
the Wigner Eq.~(\ref{eq:c-number-Wigner-dimer}).

\section{Numerical simulations}
\label{sec:num}

The choice of using the Wigner representation for the numerical
simulations is not immediately apparent, since it involves an
approximation that is not necessary if the $P$ or the $Q$-representation
are used. The advantage of the truncated Wigner Langevin
equations~(\ref{eq:langevin-W}) is that the noise is additive, as
opposed to the the multiplicative noise of the $Q$-representation (where
the noise for the quantum SHG model poses serious limits on the
parameter space \cite{bache:2002b}) and the $P$-representation. For the
$P$-representation we are forced to use the generalized representations
in order to avoid negative diffusion in the Fokker-Planck equation.
Since this choice implies doubling of the phase space, the $c$-numbers
$\alpha_j$ and $\alpha_j^\dagger$ are no longer each others complex
conjugate (only on average) and the respective noise terms are not
correlated to each other, cf.  Eq.~(\ref{eq:noise-P}). This may lead
to divergent trajectories where the convergence is extremely slow, and
is the major reason for us avoiding a numerical implementation of the
positive-$P$ equations. The Wigner equations, on the other hand, have no
problems in this direction.

The drawbacks to using the Wigner equations are first of all that we
have to neglect the third order terms of the Wigner QPD
equation~(\ref{eq:c-number-Wigner-dimer}) to get it on Fokker-Planck
form so the equivalent Langevin
equations~(\ref{eq:lang-ito-general-simple}) may be obtained. It is
uncertain what the implications of this approximation are, however in
many cases no major differences have been observed between simulations
of the truncated Wigner equations compared to exact positive-$P$ or
$Q$ equations \cite{olsen:1999,bache:2002b}. On the other hand in
Ref.\cite{kinsler:1991} so-called quantum jump processes in the
degenerate OPO above threshold are shown to produce significant
differences between the truncated Wigner and the positive
$P$-representation. In our case the third order terms are
$O(\kappa^4)$ while the other terms are $O(\kappa^2)$ or lower. And
because of the weak nonlinear coupling the effect of the third order
terms is weak, justifying the truncation. Another drawback to
the Wigner representation is that the intracavity averages are
symmetrically ordered, and these cannot be rewritten to time and
normal ordering since the intracavity commutator relations are not
known for $t\neq t'$ (only the output fields have well defined
correlations here). This means that in order to compute the output
fields at a given time $t$ we are forced to use Eq.~(\ref{eq:aout}),
and here the Gaussian white noise part of the input field is an
ill-defined instantaneous quantity. The output fields of the numerics
are calculated by using the fact that although the derivative and
instantaneous value of a stochastic term are ill-defined, the integral
is well-defined. By integrating over a time window (which we denote
$\Delta\tau$) and calculating the average, as described in
Ref.~\cite{zambrini:2000}, we may obtain the output fields from
Eq.~(\ref{eq:aout}).

We use the Heun method \cite{sanmiguel:2000} to numerically solve the
Langevin equations for the intracavity fields and to evaluate the
output fields, and a random number generator \cite{toral:1993} for
generating the Gaussian noise terms. The time step was set to $\Delta
t=0.001$ and checked to be stable. The size of the time window for
calculating the output fields was varied between $\Delta\tau= 40\Delta
t-200\Delta t$ according to the resolution needed for the individual
spectra. Finally, we set the noise strength parameter $n_s=10^8$,
which is a typical value for the cavity configuration considered here
\cite{unpublished}.

The averages calculated using the output fields of the numerics
correspond to symmetrically ordered averages since they are calculated
from the Wigner Langevin equations. In order to relate these averages
to the normally ordered averages of the spectra in
Sec.~\ref{sec:squeezing-analysis}, the output commutator
relations~(\ref{eq:out-comm}) are used to rewrite the output
correlations. The classical steady states of the output fields are
found from the average of Eq.~(\ref{eq:aout}) as
\begin{mathletters}
  \label{eq:F_out}
\begin{eqnarray}
{\mathcal F}_{1,{\mathrm out}}&=&\sqrt{2\gamma_1}{\mathcal
  F}_1+{\mathcal E_{\mathrm p}}/\sqrt{2\gamma_1} 
\\
{\mathcal F}_{2,{\mathrm out}}&=&\sqrt{2\gamma_2}{\mathcal F}_2, \quad
{\mathcal F =A,B},
\end{eqnarray}
\end{mathletters}
by taking the input fluctuation to be zero on average. Assuming that
the output fields are fluctuating around the output steady states
\begin{equation}
  \label{eq:deltaA}
  \Delta \hat A_{j,{\mathrm out}}=\hat A_{j,{\mathrm
  out}}-\mA_{j,{\mathrm out}} ,
\end{equation}
we introduce the photon number fluctuation operator for
waveguide A as
\begin{eqnarray}
    \Delta \hat N_{j,{\mathrm out}}^A(t)&\equiv& \hat N_{j,{\mathrm
    out}}^A(t)- \la \hat N_{j,{\mathrm out}}^A\ra_s 
\nonumber\\
&\simeq&\mA_{j,{\mathrm out}} \Delta \hat A_{j,{\mathrm out}}^\dagger(t)
+\mA_{j,{\mathrm out}}^* \Delta \hat A_{j,{\mathrm out}}(t),
    \label{eq:deltaN}
\end{eqnarray}
with subscript $s$ to indicate that the average is symmetric and we
have neglected higher order terms in the fluctuations. Using this
expression, the two-time correlation function~(\ref{eq:var-tb}) is
with a symmetric ordering of the operators to leading order
\begin{eqnarray}
  \label{eq:C(tau)_s}
  C_{A_jB_k}^{(\pm)}(\tau)\simeq
\la &&\Delta \hat N^A_{j,{\mathrm out}}(t)\pm
  \Delta \hat N^B_{k,{\mathrm out}}(t), 
\nonumber\\
 && \Delta \hat N^A_{j,{\mathrm out}}(t+\tau)\pm \Delta \hat N^B_{k,{\mathrm
      out}}(t+\tau) \ra_s.
\end{eqnarray}

The symmetric $c$-number correlations of the output field fluctuations
are now calculated in the numerical simulations of the dimensionless
Wigner Langevin equations~(\ref{eq:langevin-W}) as $\la \Delta
{\mathbf w}^s(0),[\Delta {\mathbf w}^s(\tau)]^T\ra_s$ where
\begin{eqnarray}
\nonumber
  \Delta{\mathbf w}^s=
\left[
\Delta {A_1}, \Delta A_1^*, \Delta {A_2}, \Delta
A_2^*, \Delta {B_1}, \Delta B_1^*, \Delta {B_2},
\Delta B_2^*
\right]^T.
\end{eqnarray} 
The spectral matrix ${\mathbf S}^s(\omega)$ of fluctuations is now
straightforwardly given by the Fourier transform of these
correlations, and the correlations~(\ref{eq:C(tau)_s}) may now be
calculated in the same manner as shown in
Sec.~\ref{sec:squeezing-analysis} with the shot-noise level
\begin{eqnarray}
\nonumber  
C_{SN}=(|{\mathcal A}_{{\mathrm out},j}|^2+|{\mathcal B}_{{\mathrm
  out},k}|^2)n_s^{-1},
\end{eqnarray}
using that the normalization of the output fields are the same as the
one taken for the input fields in Eq.~(\ref{eq:in-norm}).  Note that
the shot noise level, here expressed in the symmetric averages in the
Wigner Langevin equation, is identical to the shot-noise level
expressed in averages from the positive-$P$ equations~(\ref{eq:SN_p}),
since $|{\mathcal A}_{{\mathrm out},j}|^2=2\bar \gamma_j\bar I_j$.
This is due to the approximation made in Eq.~(\ref{eq:deltaN}).

In the analytical treatment in Sec.~\ref{sec:squeezing-analysis} we
used that in the spectral matrix ${\mathbf S}(\omega)$ certain
symmetries are present, so in fact only approximately one third of the
64 correlations were needed to obtain the results presented there. In
a numerical simulation this is only approximately valid in the limits
of long integration times and large correlation times.  Much better
results are obtained faster if the spectra are calculated directly
from the full $8\times 8$ matrix ${\mathbf S}^s(\omega)$.

\bibliographystyle{c:/LocalTexMf/miktex/prsty}
\bibliography{misc,c:/Projects/Bibtex/literature}

\begin{thebibliography}{10}

\bibitem{lugiato:1999}
L.~A. Lugiato, M. Brambilla, and A. Gatti,  in {\em Advances in Atomic,
  Molecular and Optical Physics}, edited by B. Bederson and H. Walther
  (Academic, Boston, 1999), Vol.~40, p.\ 229.

\bibitem{kimble:1987}
{\em Squeezed states of the electromagnetic field}, edited by H.~J. Kimble and
  D.~F. Walls (Special issue of J. Opt. Soc. Am. B {\bf 4}, 1449, 1987).

\bibitem{davidovich:1996}
L. Davidovich, Rev. Mod. Phys. {\bf 68},  127  (1996).

\bibitem{drummond:1981}
P.~D. Drummond, K.~J. McNeil, and D.~F. Walls, Opt. Acta {\bf 28},  211
  (1981).

\bibitem{collett:1985}
M.~J. Collett and D.~F. Walls, Phys. Rev. A {\bf 32},  2887  (1985).

\bibitem{slusher:1985}
R.~E. Slusher, L.~W. Hollberg, B. Yurke, J.~C. Mertz, and J.~F. Valley, Phys.
  Rev. Lett. {\bf 55},  2409  (1985).

\bibitem{pereira:1988}
S.~F. Pereira, M. Xiao, H.~J. Kimble, and J.~L. Hall, Phys. Rev. A {\bf 38},
  4931  (1988).

\bibitem{scott:1990}
A.~C. Scott and P.~L. Christiansen, Phys. Scripta {\bf 42},  257  (1990).

\bibitem{pettiaux:1991}
N.~P. Pettiaux, P. Mandel, and C. Fabre, Phys. Rev. Lett. {\bf 66},  1838
  (1991).

\bibitem{bache:2002b}
M. Bache, P. Scotto, R. Zambrini, M. {San Miguel}, and M. Saffman, Phys. Rev. A
  {\bf 66},  013809  (2002).

\bibitem{reynaud:1987}
S. Reynaud, C. Fabre, and E. Giacobino, J. Opt. Soc. Am. B {\bf 4},  1520
  (1987).

\bibitem{heidmann:1987}
A. Heidmann, R.~J. Horowicz, S. Reynaud, E. Giacobino, C. Fabre, and G. Camy,
  Phys. Rev. Lett. {\bf 59},  2555  (1987).

\bibitem{lugiato:1982}
L.~A. Lugiato and G. Strini, Opt. Commun. {\bf 41},  67  (1982).

\bibitem{eschmann:1999}
A. Eschmann and R.~J. Ballagh, Phys. Rev. A {\bf 60},  559  (1999).

\bibitem{fabiny:1993}
L. Fabiny, P. Colet, R. Roy, and D. Lenstra, Phys. Rev. A {\bf 47},  4287
  (1993).

\bibitem{serrat:2001}
C. Serrat, M.~C. Torrent, J. Garcia-Ojalvo, and R. Vilaseca, Phys. Rev. A {\bf
  64},  041802(R)  (2001).

\bibitem{serrat:2002}
C. Serrat, M.~C. Torrent, J. Garcia-Ojalvo, and R. Vilaseca, Phys. Rev. A {\bf
  65},  053815  (2002).

\bibitem{khoury:1999}
A.~Z. Khoury, Phys. Rev. A {\bf 60},  1610  (1999).

\bibitem{peschel:1997}
T. Peschel, U. Peschel, F. Lederer, and B.~A. Malomed, Phys. Rev. E {\bf 55},
  4730  (1997).

\bibitem{bang:1997}
O. Bang, P.~L. Christiansen, and C.~B. Clausen, Phys. Rev. E {\bf 56},  7257
  (1997).

\bibitem{perina:2000}
J. {Perina Jr} and J. Perina,  in {\em Progress in optics}, edited by E. Wolf
  (Elsevier, Holland, 2000), Vol.~41, p.\ 361.

\bibitem{wynar:2000}
R. Wynar, R. Freeland, D.~J. Han, C. Ryu, and D.~J. Heinzen, Science {\bf 287},
   1016  (2000).

\bibitem{heinzen:2000}
D.~J. Heinzen, R. Wynar, P.~D. Drummond, and K.~V. Kheruntsyan, Phys. Rev.
  Lett. {\bf 84},  5029  (2000).

\bibitem{drummond:1998}
P.~D. Drummond, K.~V. Kheruntsyan, and H. He, Phys. Rev. Lett. {\bf 81},  3055
  (1998).

\bibitem{poulsen:2001}
U.~V. Poulsen and K. M{\o}lmer, Phys. Rev. A {\bf 63},  023604  (2001).

\bibitem{milburn:1997}
G.~J. Milburn, J. Corney, E.~M. Wright, and D.~F. Walls, Phys. Rev. A {\bf 55},
   4318  (1997).

\bibitem{opo}
The model for degenerate optical parametric oscillation is found by simply
  having the classical pump field in the SH mode, so this term would be $i\hbar
  ({\mathcal E}_{{\mathrm p},O} \hat O_2^\dagger -{\mathcal E}_{{\mathrm
  p},O}^*\hat O_2)$.

\bibitem{walls:1994}
D.~F. Walls and G.~J. Milburn, {\em Quantum Optics} (Springer Verlag, Berlin,
  1994).

\bibitem{carmichael}
H.~J. Carmichael, {\em Statistical methods in quantum optics 1} (Springer
  Verlag, Berlin, 1999).

\bibitem{mandel:1995}
L. Mandel and E. Wolf, {\em Optical coherence and quantum optics} (Cambridge,
  New York, 1995).

\bibitem{ito-stratonovich}
In fact, in the present model both Ito and Stratonovich rules for interpreting
  the stochastic integration give identical results.

\bibitem{kinsler:1991}
P. Kinsler and P.~D. Drummond, Phys. Rev. A {\bf 43},  6194  (1991).

\bibitem{detuning}
If independent control over the detuning is desired a dichroic mirror could be
  inserted in the cavity, see P. Lodahl, M. Bache, and M. Saffman, Phys. Rev. A
  {\bf 63}, 023815 (2001), however this would introduce additional losses, and
  hence also noise sources, into the system.

\bibitem{saleh:1991}
B.~E.~A. Saleh and M.~C. Teich, {\em Fundamentals of photonics} (John Wiley,
  New York, 1991).

\bibitem{lodahl:1999}
P. Lodahl and M. Saffman, Phys. Rev. A {\bf 60},  3251  (1999).

\bibitem{coupling}
Alternatively, the coupling can be achieved by launching two identical modes in
  a bulk crystal. Taking the modes to be Gaussians with waists $w_j$ and a
  distance $d$ apart, the coupling coefficient can be found as an overlap
  integral of the modes \cite{fabiny:1993}. After normalization the coupling
  constants become $ \tilde J_j=\frac{2}{T_1}e^{-d^2/2w_j^2}.$ This approach is
  valid when diffraction can be neglected, which can be achieved by choosing
  $w_j$ large, typically 100 $\mu$m or more. In contrast, the width of the
  waveguides can typically be chosen to 5-10 $\mu$m. Additionally, it must be
  stressed in the bulk case that since there is no waveguiding the energy
  exchange between the spatially overlapping fields may become complicated and
  beyond the theory presented here.

\bibitem{lodahl:2001}
P. Lodahl, M. Bache, and M. Saffman, Phys. Rev. A {\bf 63},  023815  (2001).

\bibitem{etrich:1997}
C. Etrich, U. Peschel, and F. Lederer, Phys. Rev. E {\bf 56},  4803  (1997).

\bibitem{classicalP}
We consider only solutions in the classical subspace of the positive-$P$
  representation, where ${\mathcal A}_j^\dagger={\mathcal A}_j^*$.

\bibitem{gardiner:1985}
C.~W. Gardiner and M.~J. Collett, Phys. Rev. A {\bf 31},  3761  (1985).

\bibitem{reid:1988}
M.~D. Reid, Phys. Rev. A {\bf 37},  4792  (1988).

\bibitem{olsen:1999}
M.~K. Olsen, S.~C.~G. Granja, and R.~J. Horowicz, Opt. Commun. {\bf 165},  293
  (1999).

\bibitem{etrich:1997b}
C. Etrich, U. Peschel, and F. Lederer, Phys. Rev. Lett. {\bf 79},  2454
  (1997).

\bibitem{lugiato:1995}
L. Lugiato and G. Grynberg, Europhys. Lett. {\bf 29},  675  (1995).

\bibitem{olsen:2000}
M.~K. Olsen, R.~J. Horowicz, L.~I. Plimak, N. Treps, and C. Fabre, Phys. Rev. A
  {\bf 61},  021803  (2000).

\bibitem{drummond:1980a}
P.~D. Drummond and C.~W. Gardiner, J. Phys. A {\bf 13},  2353  (1980).

\bibitem{yuen:1986}
H.~P. Yuen and P. Tombesi, Opt. Commun. {\bf 59},  155  (1986).

\bibitem{gilchrist:1997}
A. Gilchrist, C.~W. Gardiner, and P.~D. Drummond, Phys. Rev. A {\bf 55},  3014
  (1997).

\bibitem{zambrini:2000}
R. Zambrini, M. Hoyuelos, A. Gatti, P. Colet, L. Lugiato, and M. {San Miguel},
  Phys. Rev. A {\bf 62},  063801  (2000).

\bibitem{sanmiguel:2000}
M. {San Miguel} and R. Toral,  in {\em Instabilities and Nonequilibrium
  Structures VI}, {\em Nonlinear phenomena and complex systems}, edited by E.
  Tirapegui, J. Martinez, and R. Tiemann (Kluwer Academic Publishers,
  Dordrecht, 2000), p.\ 35.

\bibitem{toral:1993}
R. Toral and A. Chakrabarti, Comput. Phys. Commun. {\bf 74},  327  (1993).

\bibitem{unpublished}
M. Bache, unpublished, (2001).

\end{thebibliography}

\end{multicols}
\end{document}